\documentclass[pre,aps,twocolumn,showpacs,floatfix,10pt,superscriptaddress]{revtex4-1}
\usepackage{graphicx}% Include figure files
\usepackage{dcolumn}% Align table columns on decimal point
\usepackage{bm}% bold math
\usepackage{subfigure}
\usepackage{epsfig}
\usepackage{afterpage}
\usepackage[colorinlistoftodos]{todonotes}
\usepackage[colorlinks=true, allcolors=blue]{hyperref}
\newcommand{\be}{\begin{equation}}
\newcommand{\ee}{\end{equation}}
\newcommand{\bea}{\begin{eqnarray}}
\newcommand{\eea}{\end{eqnarray}}
\newcommand{\beq}{\begin{equation}}
\newcommand{\eeq}{\end{equation}}
\newcommand{\beqa}{\begin{eqnarray}}
\newcommand{\eeqa}{\end{eqnarray}}

\begin{document}
	
	\preprint{AIP/123-QED}
	
	\title[Hierarchical Models of Water]{Hierarchical Lattice Models of Hydrogen Bond Networks in Water}% Force line breaks with \\
	%\thanks{Footnote to title of article.}
	
	\author{Rahul Dandekar}
	\email{rsdandekar@imsc.res.in}
	\altaffiliation[Also at ]{Institute of Mathematical Sciences - HBNI, CIT Campus, Taramani, Chennai, India}%Lines break automatically or can be forced with \\
	\author{Ali A. Hassanali}%
	\email{ahassana@ictp.it}
	\affiliation{ 
	Condensed Matter and Statistical Physics Section \\
        The International Center for Theoretical Physics, \\
        Strada Costiera 11, Trieste, Italy
	}%
	
	\date{\today}% It is always \today, today,
	
	\begin{abstract}
	We develop a graph-based model of the hydrogen bond network in water, with a view towards quantitatively modeling the molecular-level correlational structure of the network. The networks are formed are studied by the constructing the model on two infinite-dimensional lattices. Our models are built \emph{bottom up}, based on microscopic information coming from atomistic simulations, and we show that the predictions of the model are consistent with known results from ab-initio simulations of liquid water. We show that simple entropic models can predict the correlations and clustering of local-coordination defects around tetrahedral waters observed in the atomistic simulations. We also find that orientational correlations between bonds are longer ranged than density correlations, and determine the directional correlations within closed loops and show that the patterns of water wires within these structures are also consistent with previous atomistic simulations. Our models show the existence of density and compressibility anomalies, as seen in the real liquid, and the phase diagram of these models is consistent with the singularity-free scenario previously proposed by Sastry and co-workers (Sastry et al, PRE 53, 6144 (1996)).
	\end{abstract}
	
	\pacs{Valid PACS appear here}% PACS, the Physics and Astronomy
	\keywords{Suggested keywords}%Use showkeys class option if keyword
	\maketitle

\section{Introduction}

Water is a universal solvent for a wide range of different solutes such as ions\cite{marcus2009}, biological systems such as proteins and DNA\cite{ball2008,funel2016} and is also found near extended interfaces relevant for many different physical and chemical processes\cite{waterinterface2016}. Unlike other simple liquids, water features numerous anomalies such as the existence of a temperature of maximum density and also a very high heat capacity and dielectric constant\cite{Nilsson2015}. Despite long study from both experimental and theoretical fronts, understanding the microscopic origins of these anomalies continues to be a contentious area of research\cite{gallo2016water,debenedetti2003supercooled,angel2014nature,pettersson2016}.Over the last decade, there have been been experiments challenging the extent of the local tetrahedrality of water\cite{tokushima2008high,wernet2004structure}. \\

The predominant explanation for the anomalous properties of water is the two-critical-point scenario \cite{poole1992phase,poole1994effect,harrington1997liquid,mishima1998relationship}, which posits the existence of two supercooled liquid phases, a low density liquid (LDL) and a high density liquid (HDL) separated by a first order line which terminates in a critical point. The LDL and HDL phases differ in their local orientational ordering of water molecules. The molecular structure of the LDL phase as seen in simulations, is a more open and tetrahedral structure thus occupying more volume. On the other hand, the HDL liquid, is a distorted tetrahedral structure characterized by higher entropy\cite{borick1995lattice}. Since experimentally proving the two-critical-point scenario remains a challenge \cite{perakis2017diffusive}, the field has been mostly driven by theoretical and computational studies\cite{gallo2016water}.\\

Besides atomistic based simulations \cite{palmer2013liquid,liu2012liquid,limmer2011putative}, there have also been numerous theoretical studies using simplified lattice models to study particular properties of water. The simplest models in this regard are coarse-grained models, such as three-state models \cite{ciach2008simple}, where each lattice site represents a mesoscopic region of the liquid. In most other models, each lattice site represents a single water model, with the state of the bonds representing the presence or absence of Hydrogen bonds. Models like the Bell \cite{bell1972statistical,bell1976three,meijer1981phase} and Besseling-Lykema \cite{besseling1994equilibrium} models consider water molecules to be always constrained to be tetrahedral molecules with orientational interactions between nearest neighbours. This results in two competing orientational orderings at low-temperatures which subsequently results in two phases which can be interpreted to be the HDL and LDL phases. Other models have also been developed where interesting physics results from the interplay of tetrahedral bonding with density \cite{henriques2005liquid}. It is also worth mentioning the Mercedes-Benz model as an example of an off-lattice model which takes into account orientational interactions, but only considers tetrahedral molecules \cite{bizjak2007three,bizjak2009theory}. It has been used to model the hydrophobic interaction between non-polar solutes and water \cite{urbic2017analytical}.\\

An interesting class of models which allows for non-tetrahedrality was studied by Stanley, Sastry, Franzese and co-workers \cite{borick1995lattice,sastry1993limits,sastry1996singularity,franzese2002liquid,stokely2010effect}. These models differ from the previous ones because the HDL and LDL phases vary in their local tetrahedrality. The density and compressibility anomalies result from assigning different volumes to tetrahedral and non-tetrahedral configurations. The model developed in this paper is related to this class of models, and we discuss these models further when we investigate the phase diagram for our model in section V.\\

In most lattice models of water developed up to this point, a realistic description of the local coordination environment of the water molecules, like that described earlier, is not taken into account. Since most of these models are not built on any input from microscopic information, a comparison between the reality of supercooled water and the symmetry structure of the phases seen in the models is challenging to justify. A recent examination of some of these models \cite{pretti2009revisiting} has in fact shown that the proposed low-temperature phases obtained using these models actually correspond to amorphous solids or glasses \cite{pagnani2010discrete} rather than liquids.\\

We aim to develop models of the H-bond network that predict microscopic properties like the correlation functions and properties of small rings or loops. A possible strategy to construct such models is to make use of recent atomistic simulation data where it has been possible to study properties of the network structure\cite{Hassanali2012,Paesani2016}, and to try to extract the most relevant sets of parameters as input. In this work, we use such information from the study by Gasparotto et al. \cite{gasparotto2016probing} to construct a lattice model of water where coordination defects are included in the network and then use that to understand both the type of phase diagram of water that emerges as well as the properties of the network. Gasporotto and co-workers\cite{gasparotto2016probing} used atomistic molecular dynamics simulations to examine coordination defects in water at various temperatures as well as the structural correlations between the defects. They found that at ambient conditions, about 60\% of the water molecules accept and donate 2 hydrogen bonds. There is however, an appreciable population of both undercoordinated and overcoordinated defects, that is, water molecules that accept or donate 1 or 3 hydrogen bonds. The defects tend to cluster with each other and are characterized by a specific temperature dependence. Most of the qualitative features observed were found to be independent of the water model used, suggesting in fact that these features likely emerge from some generic properties of the hydrogen bond network. \\

We build our models from such knowledge of the population densities of the co-ordination defects in the network. Although there have been studies on the energetics of molecules with multiple H-bonds and the effect of different local configurations on the energy of a single molecule \cite{znamenskiy2007quantum,huvs2012strength}, we opt instead for a semi-empirical approach, and set the weights in our model from the densities of various types of molecules observed in simulations (as will be elaborated in section II). We study our network models analytically on two different infinite dimensional lattices, namely the 6-coordinated Bethe and Husimi lattices\cite{gujrati1995bethe}, which function as \emph{substrates} for forming the network we wish to describe. The phase diagram for our network model consists of two phases with a symmetry breaking transition between the phases, reinforcing the notion of the singularity free-scenario proposed by Sastry and co-workers\cite{sastry1996singularity}. \\

Our network models successfully reproduce several properties of the short-ranged correlations between coordination defects of the network at ambient conditions. An important result of our findings is that many of these correlations are purely entropically driven since our models do not have any explicit enthalpic interations between water molecules. We also examine the changes in these properties as a function of both temperature and pressure. Orientational correlations between bonds are also computed on the Bethe and Husimi lattices and found to be longer-range than those where orientations are neglected. In addition, topological properties like directional correlations within closed loops are also in good agreement with previous atomistic simulations \cite{bergman2000topological,hassanali2013proton}. Finally we show that the inclusion of topological defects in the model does not qualitatively change the physics of observing the anomalies like the density maximum or compressibility minimum but instead can shift the exact location of their position on the temperature-pressure phase diagram.\\

Although this will not be tackled in this work, it is worth mentioning that the model we present here serves as an important starting point for examining dynamics of the network. There have
been several atomistic based simulations showing the importance of collective network fluctuations where the reorganization of water defects, rings and wires are suggested to play an important role in dynamics\cite{ohmine1992,ohmine1996,ohmine1999}. Since the lattice model we develop here captures some of the essential physics of water networks observed in atomistic simulations, we believe that it provides a framework to examine network dynamics. This will be the subject of a forthcoming study \cite{RakalaDandekarHassanaliTobeSub}. \\

The structure of the paper is as follows: In Section 2, we detail the construction of our model and method we use to solve it on the Bethe and Husimi lattices. In Section 3, we compare the predictions of our model for more microscopic properties such as the radial distribution functions and loop statistics with data from simulations. In Section 4, we show that our model displays anomalous behaviour of the density and compressibility. In Section 5, we examine the phase diagram of the network and show that it is consistent with the singularity-free scenario. We conclude in Section 6 with a look towards future applications and extensions of our model. 

\section{Construction of the Model} \label{sec:models}

In a recent study, Gasparotto and co-workers \cite{gasparotto2016probing} used atomistic molecular dynamics simulations to probe the structural correlations between water molecules in the hydrogen bond network. In particular, they found that at ambient temperatures, about two-thirds of water molecules donate and accept two hydrogen bonds (HB). Besides these, there is also a sizable concentration of \emph{defects} in the network, such as water molecules that accept two HB and donate only one or three HB, as well as those that accept one HB and donate 1 to 3 HB. By examining the pair-correlation between the different types of water molecules, these authors showed that there are specific structural correlations between defects manifested in their tendency to cluster with each other. Interestingly, the qualitative features were found to be independent of the choice of water model. \\

As indicated earlier, most lattice models of water neglect the directionality of the hydrogen bonds that form the network, and furthermore, do not account for the existence of different types of defects such as non-tetrahedral, under- and over- coordinated water molecules. In addition, most lattice models of water have not examined the role of directed network correlations which are deemed to be important for problems involving proton transfer in water,\cite{hassanali2013proton} the reorganization of water networks around solutes like proteins\cite{rahaman2017,jong2017hydrogen}, and finally in understanding water around ions and osmolytes\cite{lee2015}. Our goal is to develop a model of a network which lives on a lattice substrate, but which encodes information on the local directionality of hydrogen bonds as well as the concentration of defects. In particular, we use the concentration of some of the important defects elucidated from the atomistic simulations as empirical parameters in our model.\\

We note here that the Bethe and Husimi lattice solutions of the model correspond to well-known approximations used to study models on finite-dimensional graphs. Our vertex model can be seen as a model for an emsemble of directed graphs with a given distribution of node in-out degrees and the solutions presented in this paper would be the mean-field solutions of the graph model. Thus one need not see the model as restricted to live on a regular lattice, or in infinite dimensions. The only spatial information incorporated into this graph model is the separate accounting of tetrahedral and non-tetrahedral 2-in-2-outs, which as we shall see is essential to producing the anomalous behaviour of water. The Bethe and Husimi solutions are then the tree-level mean field and the three-node level (Bethe-Peierls) mean field solutions\cite{gujrati1995bethe} respectively, for the directed graph ensemble described by our vertex weights. This allows for the determination of multi-site correlations self-consistently on the infinite-dimensional lattice \cite{izmailian1998exact}.

\subsection{Definition of the Vertex Weights}

\begin{figure}[t]
	\centering
	\includegraphics[width=0.9\columnwidth]{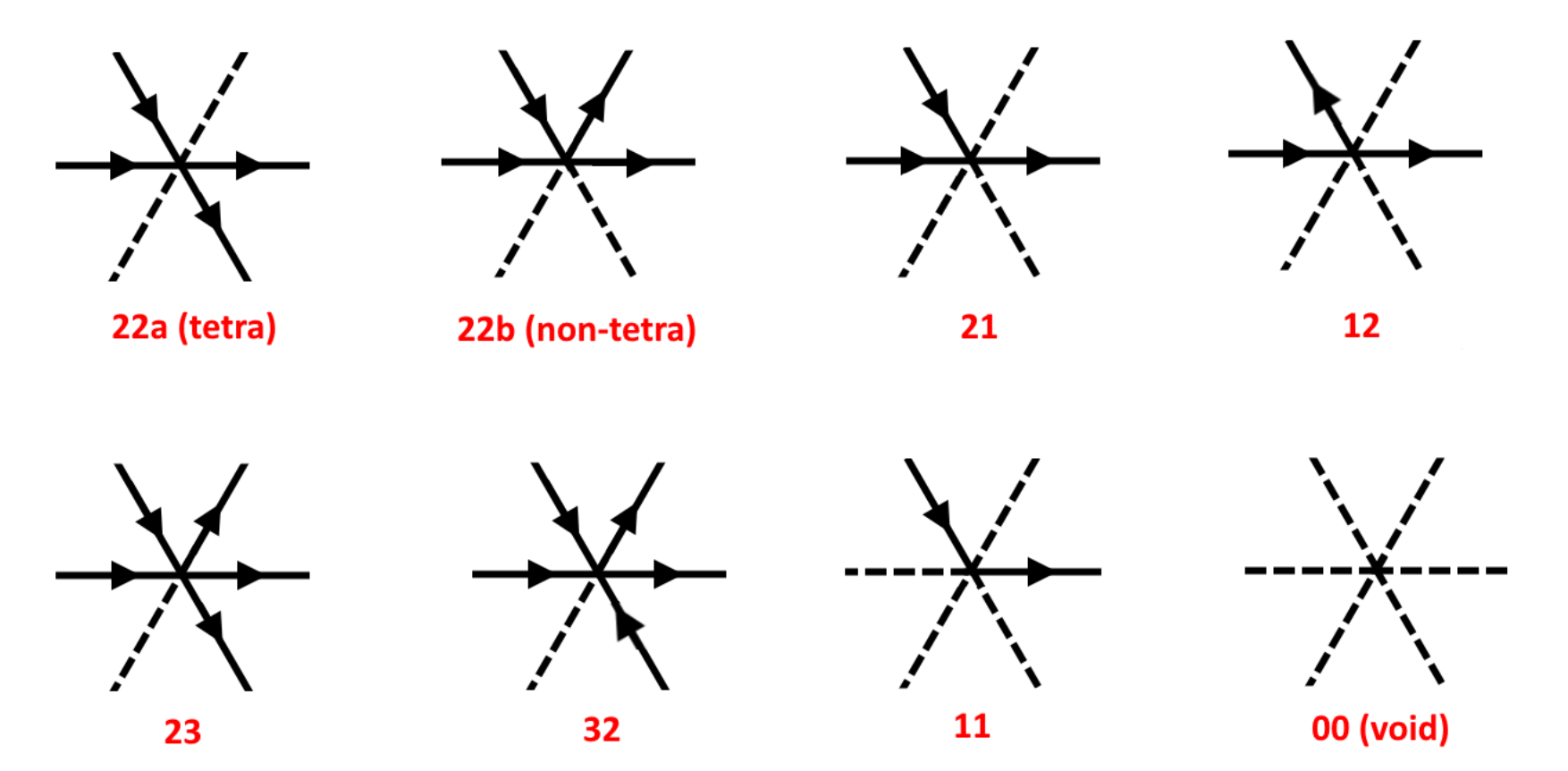}
	\caption{A sketch of the different defects that are treated in our lattice model. Labels for the defects are also shown in red (see text for more details).}
	\label{recur}
\end{figure}

Our network models are constructed on a 6 coordinated lattice with triangular faces. Although the particular details of the lattice are not important, the triangular structure is convenient as it allows for all the important defects such as over and under-coordinated water molecules to be easily included. The frustration caused by the triangular structure also destabilises crystal ordering. The sites of the lattice correspond to water molecules. Bonds on the lattice may be occupied or empty - occupied bonds correspond to H-bonds between the sites they connect, while empty bonds correspond to the \emph{interstitial} molecules within the bonding shell but not actually bonded. Besides the presence of hydrogen bonds, liquid water is also characterized by regions of empty space (cavities or voids)\cite{Chandler2005,sosso2016role}. Cavities within the liquid in our lattice model are represented by empty sites, the concentration of which can also be controlled via a separate fugacity, as has been done in previous studies with lattice models \cite{sastry1993limits}.\\

We denote the type of water molecule by the notation $xy$, where $x$ is the number of incoming hydrogen bonds and $y$ is the number of outgoing hydrogen bonds. The most common type at room temperature are molecules with 2 incoming and 2 outgoing H-bonds, denoted `22s' as observed in ab-initio simulations. The other types may be considered as `defects' in the network. In our model, we include the following types of molecules: 22, 12, 21, 32, 23, and 11, these being the most common types of defects in room temperature water as observed in ab-initio simulations. We also allow for cavities (empty sites) on the lattice with a small weight. As in some of the lattice models discussed in the previous section, one also distinguishes between tetrahedral and non-tetrahedral 22 molecules. Tetrahedral 22 molecules on our lattice are defined to be the ones where the empty sites lie opposite each other and are denoted `22a', whereas in the non-tetrahedral 22 waters, denoted `22b', the empty sites lie on the same side. The six different types of vertices are depicted in Figure~\ref{recur}. Each vertex is assigned a particular weight which is defined in the following manner: \\

\beqa
w_{`22a'} \equiv ~ w_{22}^{tetra} &=& ae\\
w_{`22b'} \equiv ~ w_{22}^{distor} &=& a\\
w_{12} = w_{21} &=& b\\
w_{32} = w_{23} &=& c\\
w_{11} &=& d \\
w_{vacancy} &=& f
\eeqa

Note that there are no explicit interactions between neighbouring lattice sites in our model. However, entropic constraints of sharing H-bonds make it more likely, for example, that a pair of neighbouring sites is 23-32 rather than 23-23.\\

For simplicity, we take the weights to be symmetric between incoming and outgoing bonds, although this is not the case in real water. Thus, our model would not capture the asymmetric properties of the directed network. As we will see later, this assumption is very accurate for the correlations of the most abundant water molecule in the network, namely, the 22 molecules. Furthermore, other network properties like the statistics of closed loops also seem to be insensitive to this assumption. The weight of a particular configuration $C$ on the lattice is given by
\beq
W(C) = \prod_i w(i)
\eeq
where $i$ runs over the vertices of the lattice.\\

There are five independent parameters in the model setting $a=1$. These parameters depend on pressure and temperature. At normal temperature and pressure, the values of three of these parameters, $b, c, d$ (relative to the weights of the 22-type molecules, $(1+e)$) such as to get the correct relative proportions of 2-, 3-, 4- and 5- co-ordinated molecules in the liquid. This method does not allow one to determine the tetrahedrality factor $e$, but we observe that the statistics of 6-rings (studied in section III C below) depend on the relative weights of tetrahedral and non-tetrahedral 22s. Based on this, we have set $e = 1.5$ on the Bethe Lattice and $e= 2.5$ on the Husimi lattice for normal temperature and pressure. Other results do not seem to depend sensitively on the precise relative weights of tetrahedral and non-tetrahedral molecules.

%For the defect correlations in the network, we stick to using thermodynamic conditions representative of room temperature and pressure (T=$300$K and P=$1$atm).

\subsection{Recursion Relations on Hierarchical Lattices}

\begin{figure}[t]
	\centering
	\subfigure[]{
		\includegraphics[width=0.43\textwidth]{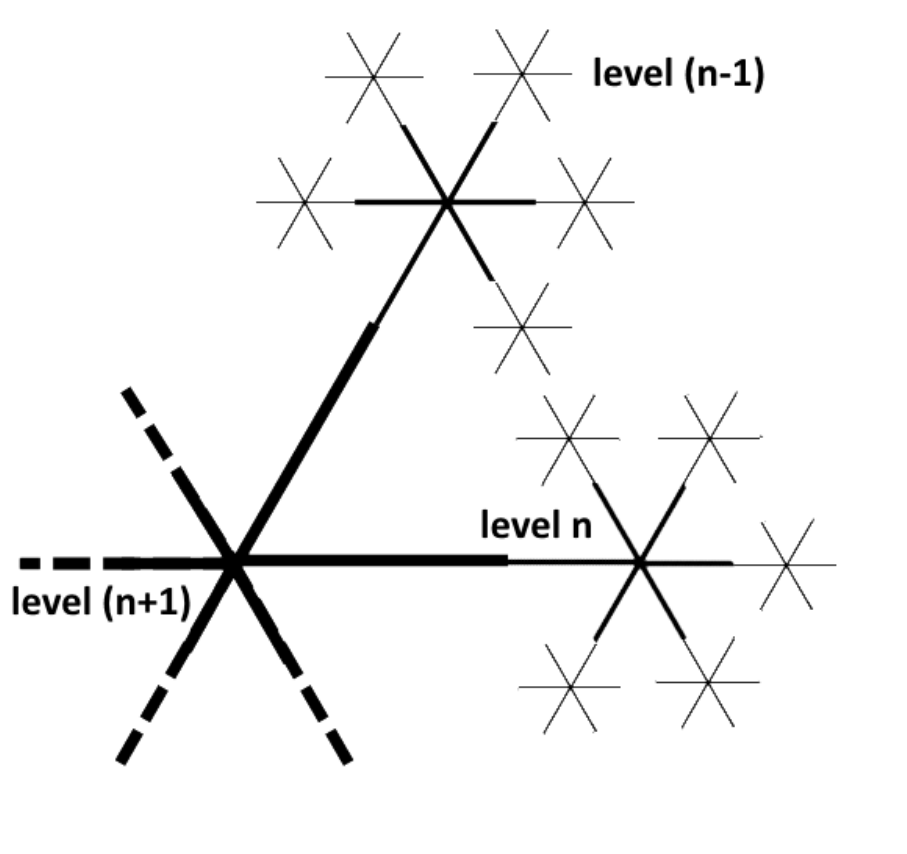}
	}
	\hspace{0.5cm}
	\subfigure[]{
		\includegraphics[width=0.4\textwidth]{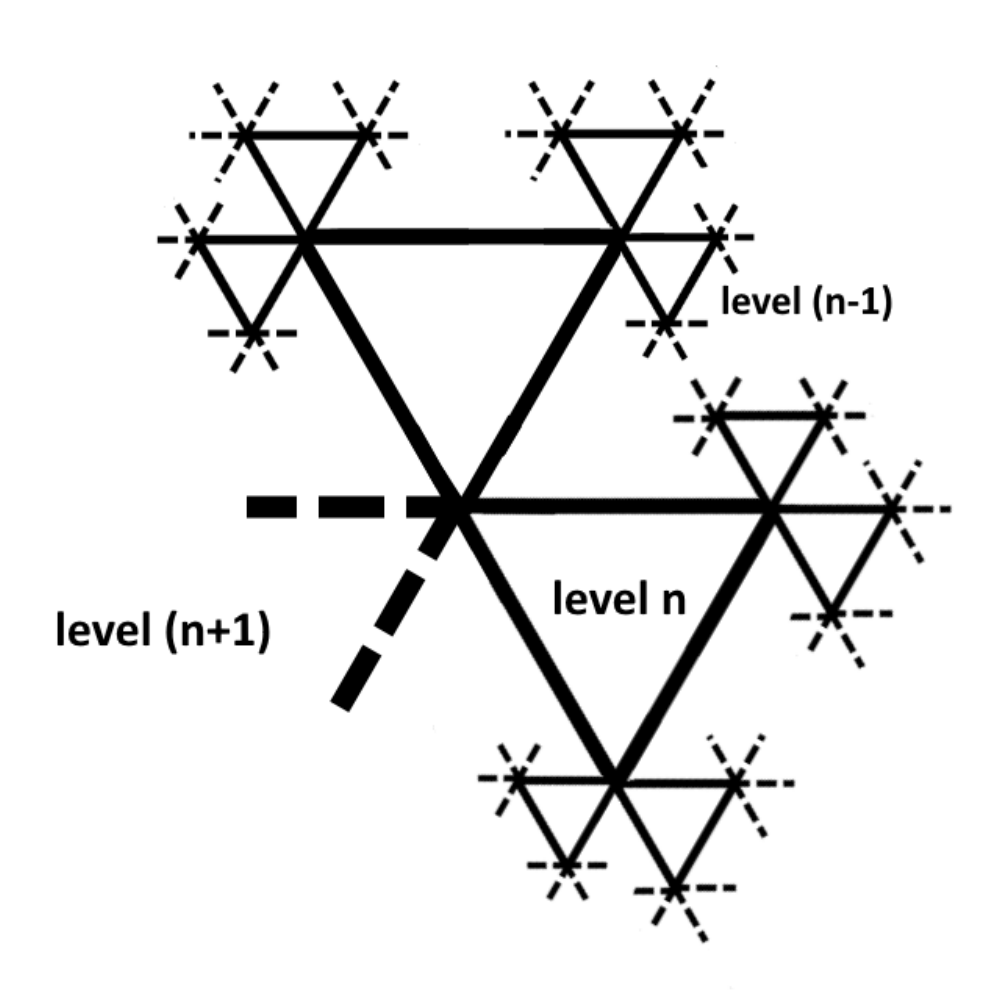}
	}
	\caption{The (a) Bethe and (b) Husimi networks}
	\label{networks}
\end{figure}

In this paper, we calculate the properties of our model on two infinite-dimensional hierarchical lattices, the Bethe and Husimi lattice generalizations of the triangular lattice, shown in Figure~\ref{networks} (a) and (b) respectively. We now begin our analysis by explaining the procedure of solving the network properties on the Bethe lattice (Figure\ref{networks} a). We begin by writing recursion relations for the restricted partition functions at level $(n+1)$ in terms of the parameters at level $n$ \cite{neto2013entropy}. One needs to find the set of conditional partition functions which are sufficient to calculate all properties on the lattice at level $n$. For the Bethe lattice in Figure~\ref{networks} a), it suffices to condition on the three orientations of the bond, and the three states of the bond (namely outgoing bond, incoming bond, and empty). Let us call the restricted partition function at level $(n+1)$ conditioned for the bond with the orientation $j$ ($=$1, 2, 3) in state $B$ as $Z^{n+1}_j(B)$. This can then be expanded as a function of the set of $\{Z^{n}_i(b)\}$ at level $n$,
\beq
Z^{n+1}_j(B) = \sum_{C'} W(C') = f(\{Z^{n}_i(b)\}) \label{eq:recurfp}
\eeq
where the sum over $C'$ includes only the configurations where the bond at level $(n+1)$ of orientation $i$ is in state $B$. An explicit example of the function $f$ for the Bethe lattice is given in eqn. (\ref{eq:recur}) below.\\

An observable at level $(n+1)$ can then be calculated from the knowledge of $\{Z^{n}_i(b)\}$. Deep inside the lattice when $n$ is large, the values for the observables converge to constant values, which are then the mean-field equilibrium predictions for the observables. On the Bethe lattice, to know the weight of a given set of local observables on a finite set of sites, it suffices to know the probabilities of the bonds connecting the set of sites to the rest of the lattice. For this reason, the set of nine parameters $\{P^n_i(b)\}$, where $P^n_i(b)$ is the probability of finding a bond at level $n$ of orientation $i$ in the state $b$, suffices to calculate all correlation functions. Deep inside the lattice, the fixed point of the recursions governing these parameters is enough to calculate all correlation functions.\\

In fact, the simple structure of the Bethe lattice allows one to calculate these properties explicitly, by counting. For example, in the normal liquid phase, knowing the probability of finding an empty bond, $r$, for a given temperature and pressure, the probability of finding a site of type $32$ is $6*{{5}\choose{2}} r(1-2r)^5$, which is the probability of finding one bond empty, three bonds pointing out (probability $(1-2r)$), and two bonds pointing in (probability $(1-2r)$, multiplied by the number of ways of choosing the empty bond (6), and then choosing two out of the five remaining bonds to be outgoing. Similarly, the probability for a pair of neighbours connected by an empty bond to be types $ab$ and $cd$ is given by (following similar reasoning) 
\beqa
P(ab,cd) = {{5}\choose{a}} {{5-a}\choose{b}} {{5}\choose{c}} {{5-c}\choose{d}} \nonumber \\ \times (1-2r)^{a+b+c+d} r^{10-a-b-c-d}
\eeqa
For this reason, we refer to the Bethe network model as the 'simplest entropic network' for our model. \\

In terms of the restricted partition functions defined earlier, $P^n_i(B) = Z^n_i(B)/\sum_b Z^n_i(b)$. At high temperatures, there is no symmetry breaking in the network, and thus, in the high temperature phase, only three parameters suffice to describe the state of the system, $P^n(b) \equiv P^n_1(b) = P^n_2(b) = P^n_3(b)$. Due to the symmetry between incoming and outgoing bonds, $P^n(+)=P^n(-)=p_n$, and $P^n(\phi)=r_n=1-2p_n$ (where $\phi$ denotes an empty bond). In the symmetric phase, the recursion equation for $r_n$ is (where $p_n = (1-r_n)/2$):

\beqa
r_{n+1} = 6 a e p_n^4 r_n + 24 a p_n^4 r_n + 30 b p_n^3 r_n^2 \nonumber \\ + 10 c p_n^5 + 20 d p_n^2 r_n^3 + f r_n^5 \label{eq:recur}
\eeqa

For the symmetry-broken low-temperature phase, one needs to analyse 3 interdependent recursion equations, one for each bond orientation.\\

The fixed point of the full recursion relations shown in Equation~\ref{eq:recurfp} gives the bulk behaviour of the network. At low temperatures, The fixed point of $r$ given by the recursion Equation~\ref{eq:recur} becomes unstable to orientation perturbations, and the symmetric phase becomes metastable. Three new stable fixed points develop, each breaking the orientational symmetry in one of three ways. In the Landau Free Energy, this corresponds to the new minima developing in the free energy surface. The phase transition occurs when the free energy of these minima is lower than the free energy of the symmetric phase. This phase transition will be explored in Section~\ref{sec:phase}. \\

The regular 6 co-ordinated Husimi network is shown in Figure~\ref{networks} b). In contrast to the Bethe lattice, which includes no loops, the Husimi lattice is a better local approximation to the network structure in water because it takes into consideration 3-loops, which correspond to clusters of 3 molecules within bonding distances of each other. We shall see that the inclusion of 3-loops improves the statistics of certain quantities, such as loops, especially on the Husimi network. The Husimi network can be solved in a similar way to the Bethe lattice, by solving for the fixed point of the minimal set of parameters. On the Husimi lattice, the minimal set of parameters is the probabilities of all states of each of the three kinds of elementary triangle, and thus is a set of $3$ x $3^3 = 81$ parameters. The set of 81 recursion relations can then be solved and the calculation of various network properties proceeds in a similar fashion to the Bethe lattice.

\subsection{Temperature and Pressure Dependence of the vertex weights} \label{sec:methods}

In order to construct the phase diagram for the lattice model and to explore the changes in network properties as a function of temperature and pressure, the weights of the lattice sites need to be modified accordingly. We choose the simplest interpolation of the dependence of the parameters of our model on temperature and pressure:
\beqa
w_i(T,P) &=& w_i^o \exp{(-( \epsilon_i + P' v_i)/T)} \label{eq:wtp}\\
&=& w_i^o \exp{(-( h_i + P v_i)/T)}
\eeqa

where the prefactor $w_i^o$, the vertex enthalpies at atmospheric pressure $h_i$ ( $= \epsilon_i + P_0 v_i$ where $\epsilon_i$ are the vertex energies, and $P_0$ is atmospheric pressure) and the local volumes $v_i$ are chosen to be independent of temperature and pressure. The $P \equiv P' - P_0$ denotes the difference of the actual pressure $P'$ from atmospheric pressure. $h_i$ are chosen such as to emulate the simulation data on the variation of the fractions of $i$-co-ordinated molecules with temperature \cite{gasparotto2016probing}. The simulations find that 4- and 5- co-ordinated molecules decrease in the same proportion with increasing temperature, while 2- and 3- co-ordinated molecules proportionately increase. For the rest of the paper we work in the variable $P$. The units of temperature and pressure are arbitrary, and we have chosen to set $T=1$ for room temperature ambient pressure to be $P=0$. \\

The local volumes $v_i$ were assigned using a strategy adopted from a previous study by Sastry et al \cite{sastry1996singularity}. These local volumes have the following functional form:
\beq
v_i = 1.0 + 0.1 n^H_i + 0.5~ \delta_{i,`22a'} + 5~ \delta_{i,void} \label{eq:vols}
\eeq

\noindent where $n_H$ is the total number of hydrogen bonds of the molecule, and the Kronecker deltas $\delta_{i,`22a'}$ and $\delta_{i,void}$ term account for the fact that locally tetrahedral configurations and voids in the fluid (considered as units) have larger volumes. The rationale behind the functional form of the local volumes is that, firstly, tetrahedral molecules occupy higher volume than non-tetrahedral molecules, secondly, the volume increases with increasing $n_H$, and lastly, the void volume should reproduce the physical observation that the density of water decreases with $T$ at high temperature. The prefactors are arbitrary and do not have any qualitative effect on the results that we report later.

%It is the increase in higher-volume locally tetrahedral configurations at low temperature %combined with the increase in the number of voids at high temperatures that gives rise to %the density anomaly in our model.

\section{Network Properties at Room Temperature} 
\label{sec:prop}

%In the previous section we showed that the phase diagram we obtain from our lattice model is consistent with earlier proposals of the singularity-free scenario.  

In the next two sections, we present results for properties of the liquid phase in our model. In this section we study the properties at room temperature and pressure, and in the next section we discuss how they vary with temperature and pressure.

In this section, we illustrate the correlations on the network, those between different bonds and well as between the different network sites, emphasizing the consistency between what we observe for the 22 molecules and the atomistic AIMD simulations. We begin with discussing our results on the structural correlations between the water molecules and defects and then also show our analysis on the orientational correlations that exist on the lattice. We also discuss the properties of different types of loops formed in our model and compare them with data from atomistic simulations.

\subsection{Structural Correlations on the Lattice}

\begin{figure}[]
	\centering
	\includegraphics[width=0.9\columnwidth]{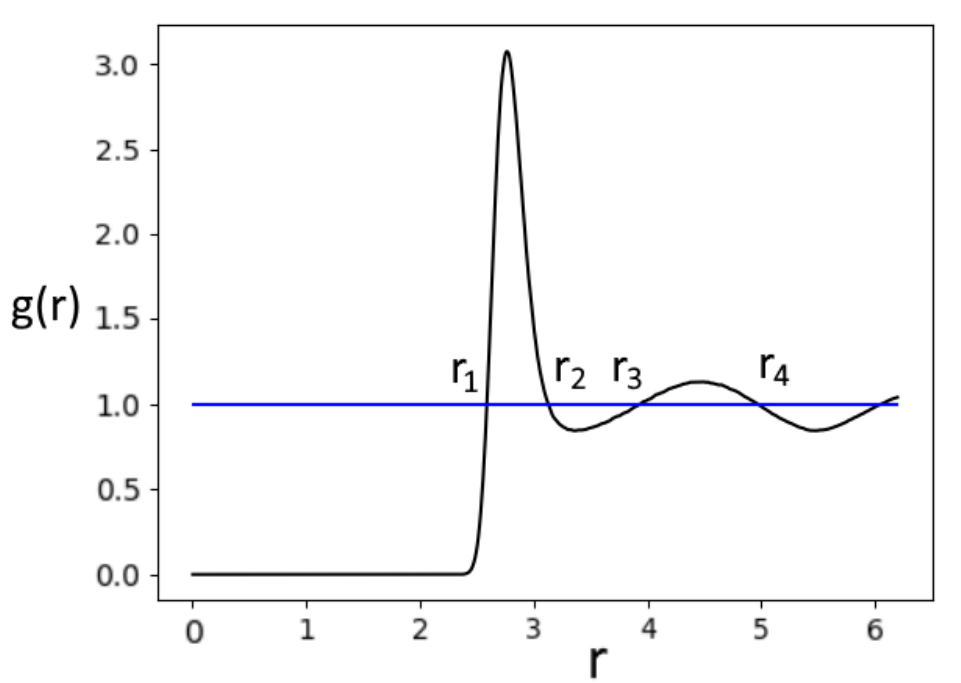}
	\caption{The water-water $g^{latt}(r)$ used to define the bonding and interstitial regions.}
	\label{shells}
\end{figure}

%\afterpage{\clearpage}

\begin{figure*}[]
	\centering
	\subfigure[]{
		\includegraphics[width=0.9\columnwidth]{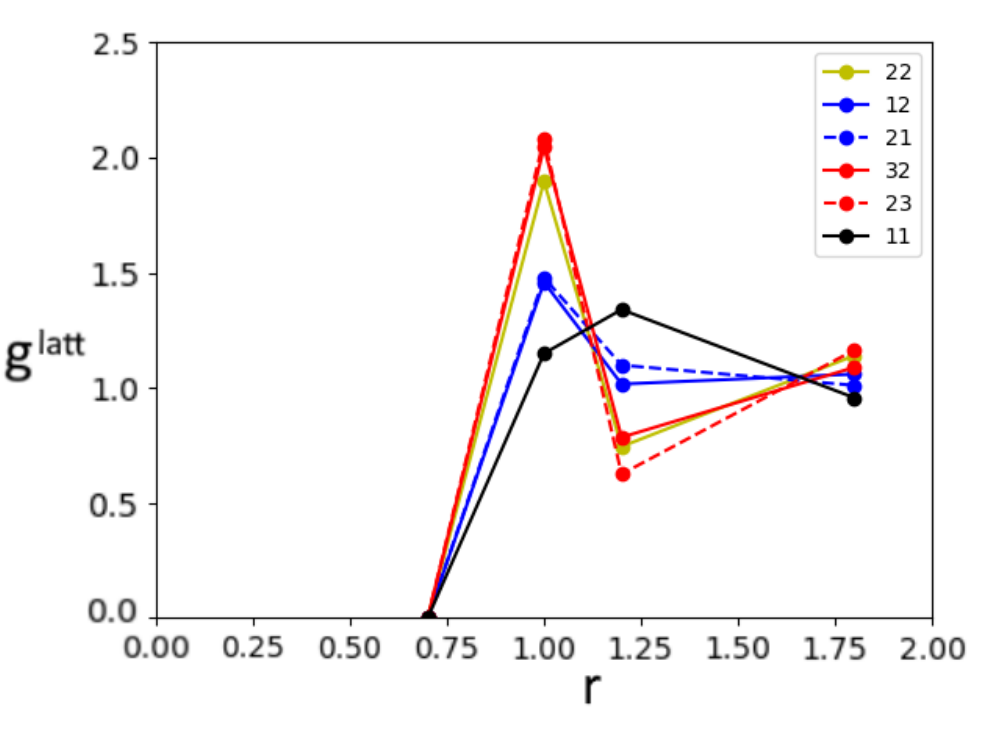}
	}
	\subfigure[]{
		\includegraphics[width=0.9\columnwidth]{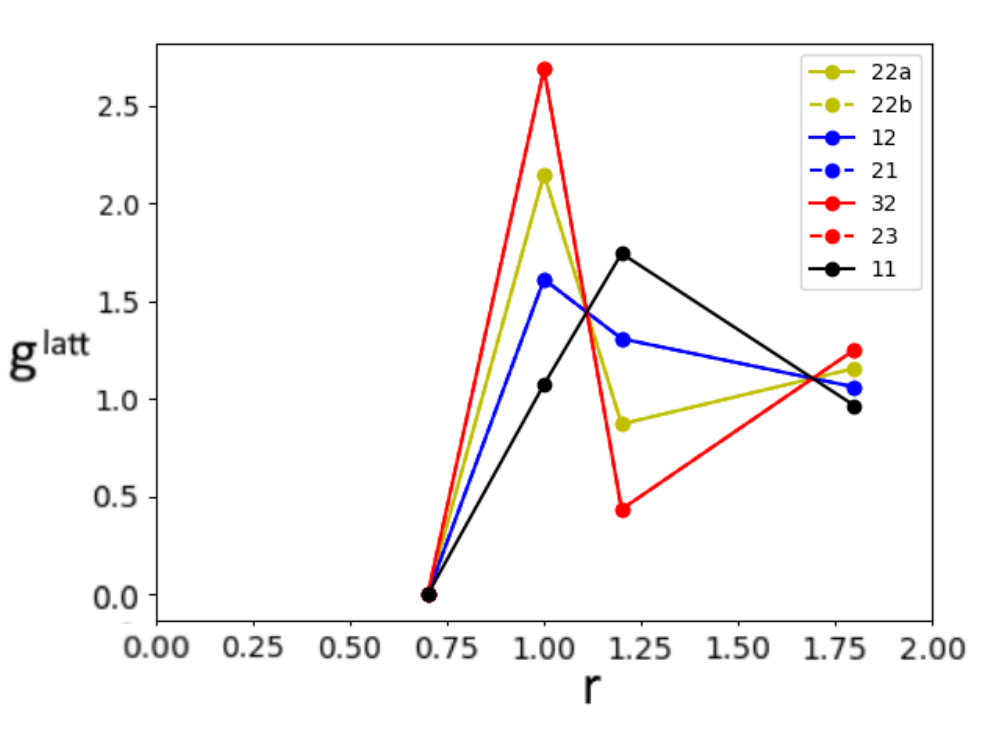}
	}
	\caption{The $g^{latt}(r)$ of various types of molecules around a generic molecule (a) from ab-initio data (b) on the Bethe lattice. The $r$-axis has been scaled such that the position of the first peak is at unit distance.}
	\label{gr_all}
\end{figure*}

\begin{figure*}[]
	\centering
	\subfigure[]{
		\includegraphics[width=0.9\columnwidth]{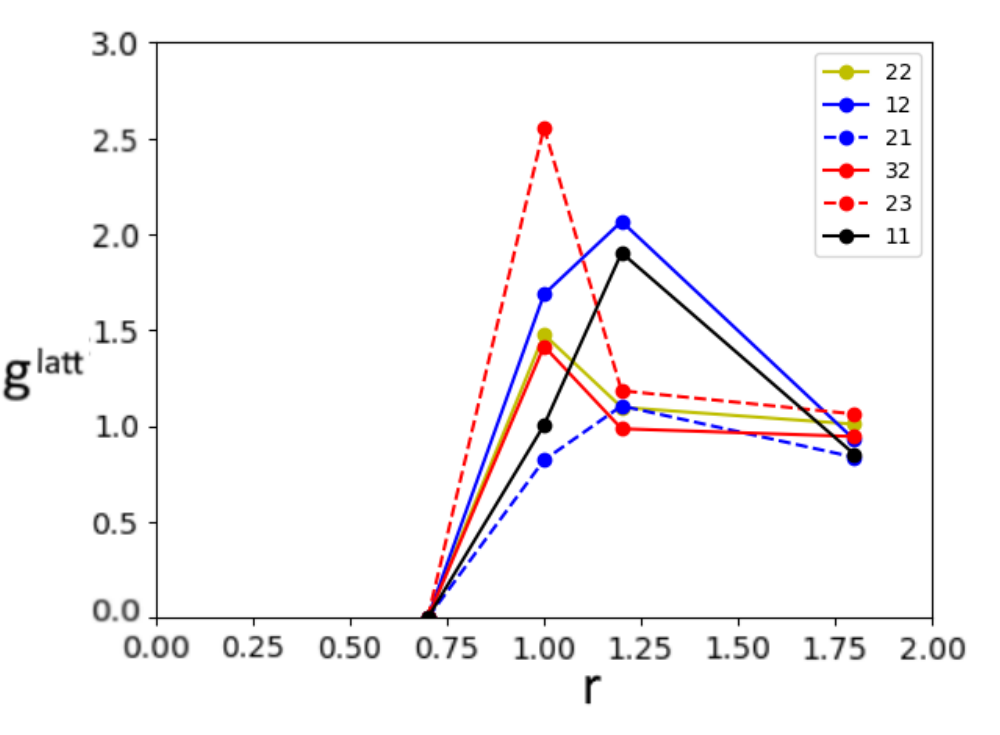}
	}
	\subfigure[]{
		\includegraphics[width=0.9\columnwidth]{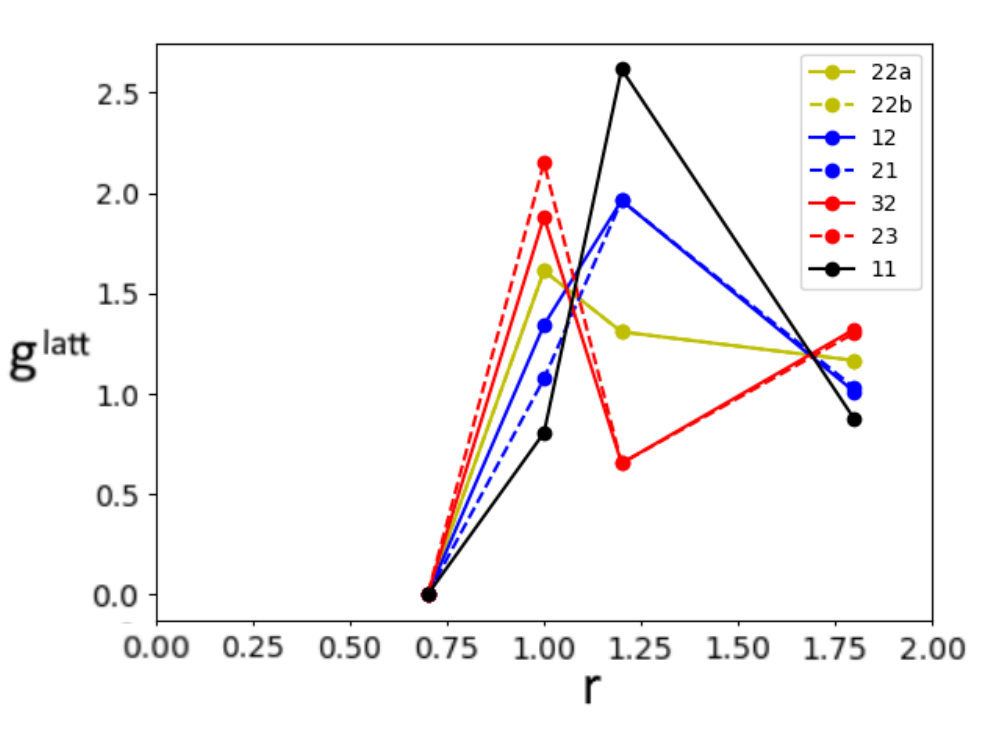}
	}
	\subfigure[]{
		\includegraphics[width=0.9\columnwidth]{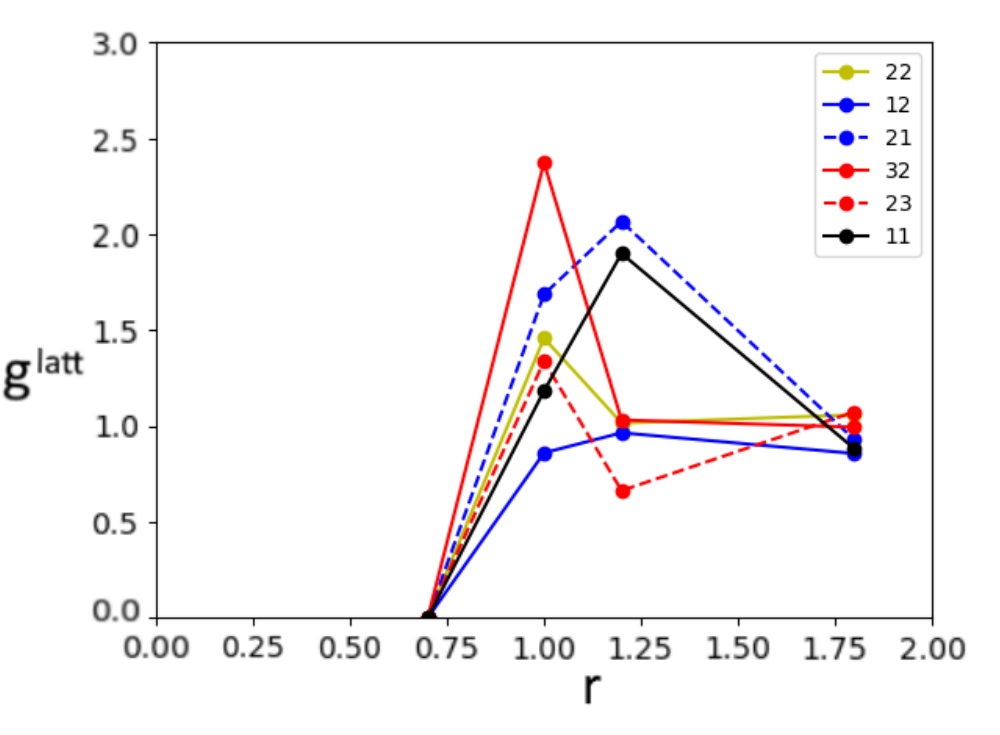}
	}
	\subfigure[]{
		\includegraphics[width=0.9\columnwidth]{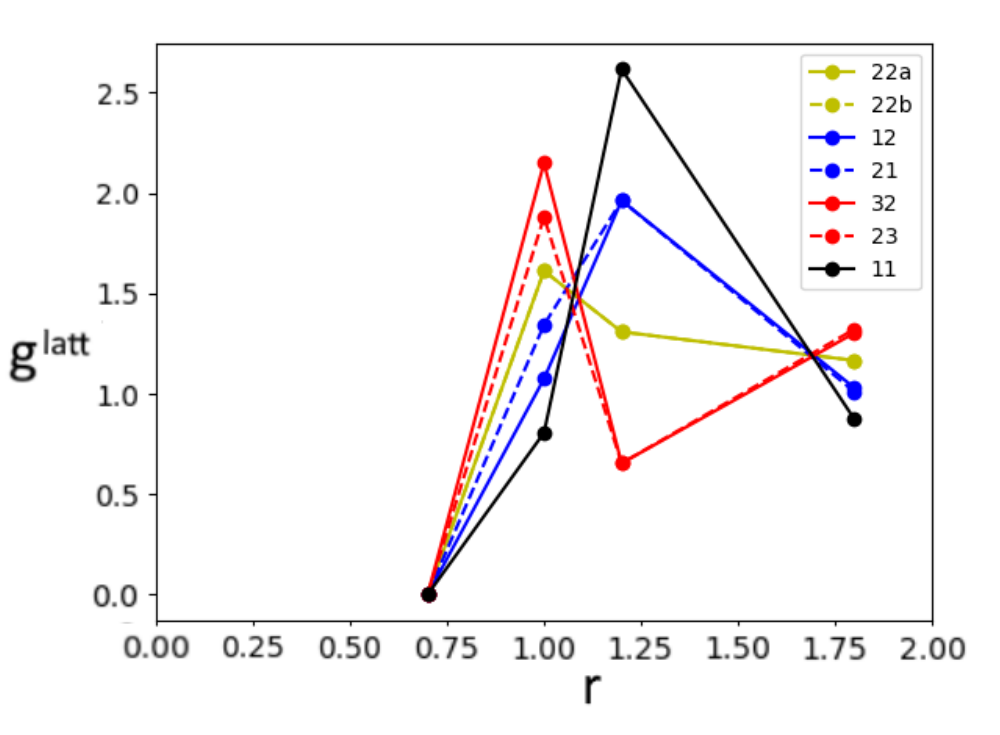}
	}
	\caption{The $g^{latt}(r)$ for various types of molecules around a $21$ molecule in (a) and (b), and a $12$ molecule in (c) and (d). (a) and (c) are plotted from ab-initio data, while (b) and (d) are results for a Bethe lattice. The $r$-axis has been scaled such that the position of the first peak is at unit distance.}
	\label{gr_21}
\end{figure*}

The correlations between different molecule types are determined by constructing radial distribution functions between different types of water molecules. The radial distribution function $g(r)$ is defined by the following equation
\beq
\int_{r_1}^{r_2} g(r) r^2 dr = \frac{1}{4\pi \rho} \langle \sum_i N^i_{r_1,r_2} \rangle
\eeq
where $N^i_{r_1,r_2}$ is the number of water molecules between the distances $r_1$ and $r_2$ from the molecule $i$, and the sum is over all the water molecules in the system. Similar radial distribution functions can be defined for specific pairs of molecule types. For example, $g_{22,11}(r)$ measures the correlations between a water molecule of type $11$ at a distance $r$ from a molecule of type $22$. An average over all types will be denoted by $*$, and thus $g_{22,*}(r)$ counts the molecules of any type at a distance $r$ from a molecule of type $22$, while $g_{*,*}(r)$ is the same as $g(r)$ defined above. Gasparotto et al \cite{gasparotto2016probing} determined the radial distribution functions for the most common type of defects in order to study the structural correlations in the network. One of the interesting and important findings of that study was that different types of defects have varying propensities to cluster with each other in the hydrogen bond network.\\

It is interesting to examine to what extent our mean-field network models are able to predict such features. A comparison between the atomistic and lattice models is also useful because it clarifies the nontrivial features in the former that are not captured by the latter, such as intermolecular interactions that are not included in the simple lattice models. In order to compare the atomistic results with our network models, we convert the atomistic radial distribution functions to information on the local network connectivity on the lattice.\\

The comparison between the atomistic data and the lattice model is achieved by finding a mapping between shells in the atomistic $g(r)$ to connectivity on the lattice. We thus integrate the atomistic $g(r)$ to get the average densities in three regions: the bonding shell, the first interstitial, and the second shell. The bonding shell is defined as the first interval $[r_1,r_2]$, starting out from the origin where $g(r)>1$. See Figure~\ref{shells} for more details. The bonding shell represents neighbouring molecules which are directly hydrogen bonded to the central molecule. Thus, on the lattice, these pairs of molecules correspond to neighbouring nodes connected by a single hydrogen bond. The first interstitial is the first interval $[r_2,r_3]$ where $g(r)<1$. On the lattice, this corresponds to neighbouring molecules which are connected to the central molecule by an empty bond. Finally, the second shell is the interval $[r_3,r_4]$ where $g(r)>1$, and these are molecules connected to the central molecule by two hydrogen bonds except those connected by two interstitials. \\

The normalized discrete $g_{a,b}(r_{i})$ in an interval $[r_i,r_{i+1}]$ is given by
\be
g_{a,b}(r_{i}) = \frac{\int_{r_i}^{r_{i+1}} g_{a,b}(r) r^2 dr}{\int_{r_i}^{r_{i+1}} r^2 dr}
\label{eq:graimd}
\ee

On the lattice, we obtain $g_{a,b}(r_{i})$ at various regimes by counting the average number of $b$ lattice sites at various states of connection to an $a$ site. On the lattice for a given state of connection, we can calculate the correlation $C_{a,b}(r)$ which measures the probability of finding an $a$ and $b$ lattice at a distance $r$ away from each other. On the lattice, $r$ of course is not a Euclidean distance but corresponds to a state of connection. This correlation is computed from knowledge of the restricted partition functions deep inside the lattice. The total number of sites at distance $r$ from an $a$-type site is denoted as $N_a(r)$. Note that this number can vary for different sites: for example, a $11$ site has four interstitial neighbours while a $22$ site has only two. We also define $N(r) \equiv \sum_a \rho_a N_a(r)$. The pair correlation function on the lattice is then given by:

\beq
g^{latt}_{a,b}(r_{i}) = \frac{N_a(r) C_{a,b}(r)}{\sum_b C_{a,b}(r) N(r) \rho_b}
\eeq

Fig. \ref{gr_all} (a) shows the reconstructed discrete radial distribution functions extracted from the atomistic simulations using Equation~\ref{eq:graimd}. In particular, we focus the discussion on $g_{22,X}$ since, as mentioned earlier, atomistic simulations indicate that the 22 sites are the most populated in the hydrogen bond network. These radial distribution functions show the tendency of different types of defects to cluster with each other. It can be seen, for example, that unlike all the other types of defects, molecules of type $11$ are most common in the interstitial region, in a state where they are not bonded with the central molecule. Interestingly, 23 and 32 sites have the largest propensity to cluster around 22 sites followed by 22 and then finally 12 and 21, in the bonding shell region. The right panel of Fig~\ref{gr_all} shows the pair correlation functions obtained from the Bethe lattice again for $g_{22,X}$. We see that the correlation functions constructed on the Bethe lattice reproduce essentially all the features that are observed from the atomistic simulations such as the enhanced population of 11 sites in the interstitial regions and the clustering of different types of defects. Thus, at least for the 22 molecules, no intermolecular interactions with other kinds of molecules need to be included in our model to reproduce the qualitative trends observed in the $g(r)$. This is an important result since it shows that the correlations observed in the atomistic simulations emerges from entropic effects. \\

%As already alluded to earlier, our lattice model assumes that the weights of the edges are symmetric between incoming and outgoing bonds and hence it cannot capture the lack of symmetry between the $23$ and $32$ defects observed in the ab initio simulations. However, it is rather interesting to see that despite its simplicity, the lattice model reproduces many of the features of the structural correlations in the pair distribution function that were observed in the atomistic simulations.\\

%\afterpage{\clearpage}

Besides the $g_{22,X}$ pair correlation functions, we also examined distributions associated with other types of defects. In Figure~\ref{gr_21} the RDFs for the 21 and 12 water molecules are shown ($g_{21,X}$ and $g_{12,X}$). While there are some trends that are reproduced by our models, the agreement in the correlations are not as favorable as those observed for $g_{22,X}$. More specifically, we see that for the 21 defects, the ordering of the 23, 12 and 11 defects in the first shell and that for the 12 defects, the ordering of the 32, 22 and 12 is reproduced by the lattice model. However, the main systematic feature not reproduced is the difference between the curves for the 12 and 21, and also that between the curves for 23 and 32 in the interstitial region. This feature is also not reproduced in the Husimi lattice $g(r)$s - refer to SI \cite{supplinfo} for details. Perhaps a part of the reason for these discrepancies is that the weights of our model are symmetric between 21 and 12, and between 23 and 32. It might also be that these asymmetrical differences are a non-trivial chemical effect which cannot be captured by a simple mean-field vertex model, and that one would have to include intermolecular interactions between the pairs 21-12, 21-21, 21-32, 21-23.\\

\begin{figure}[]
	\centering
	\includegraphics[width=0.9\columnwidth]{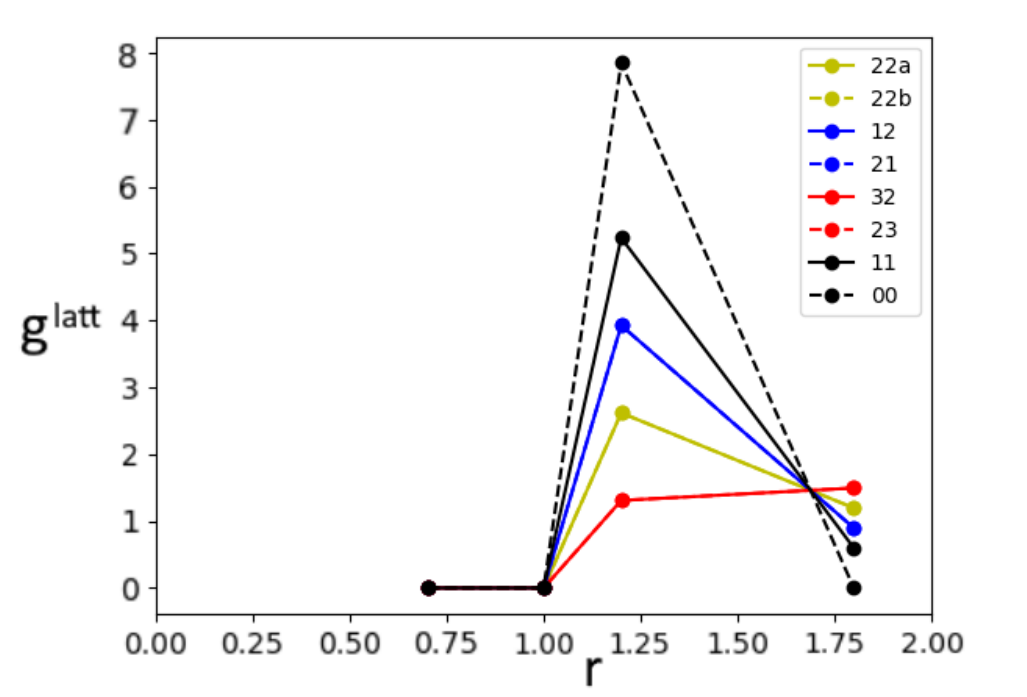}
	\caption{The $g^{latt}(r)$ of various water molecules around a void or cavity (represented on the lattice as an empty site), at normal temperature and pressure. The $r$-axis has been scaled such that the position of the first peak is at unit distance.}
	\label{voidgr}
\end{figure}

%In particular, the lattice model predicts for a 21 site, that its correlations with 12 and 21 water molecules are very similar in the interstitial region whereas in the AIMD simulations these are very different. Similarly, we see that for 23 and 32 around a 21 site, the lattice model significantly underestimates the difference observed in the AIMD calculations. Similar types of discrepancies can also be identified for the 12 sites shown in the bottom panel of Figure\ref{gr_21}.

\begin{figure*}[]
	\centering
	\subfigure[]{
		\includegraphics[width=0.9\columnwidth]{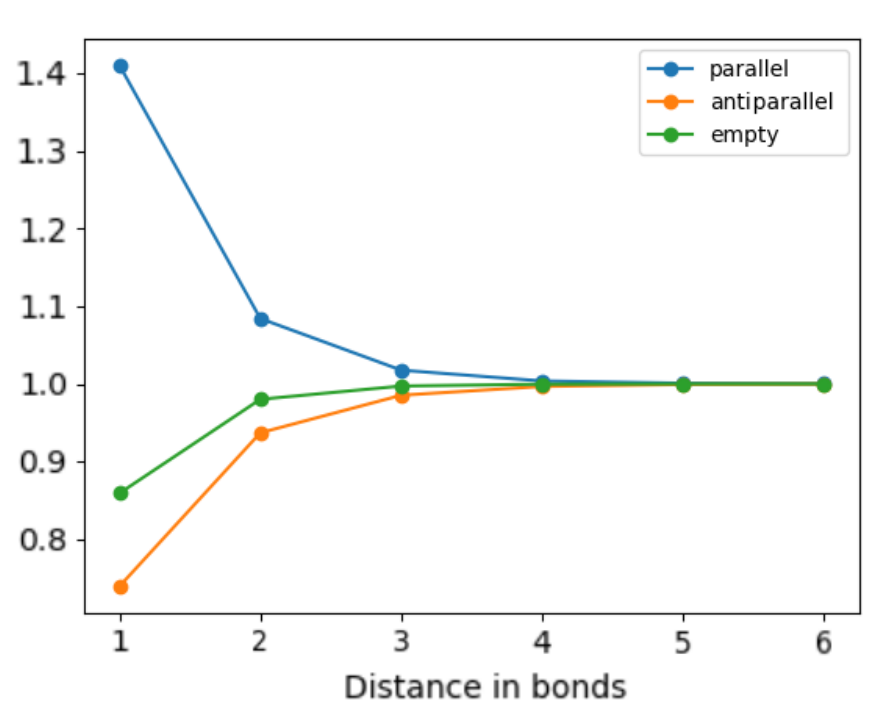}
	}
	\subfigure[]{
		\includegraphics[width=0.9\columnwidth]{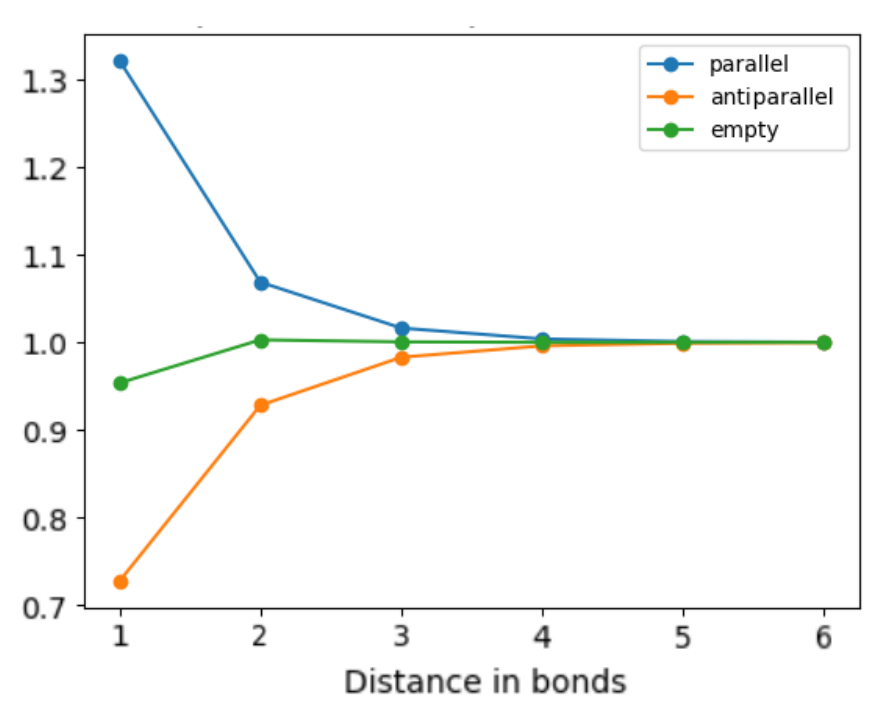}
	}
	\caption{The probability of finding a bond in a state $x$ a certain distance away from an occupied bond, relative to the equilibrium probability of finding a bond in a state $x$. Figure (a) shows the results on the Bethe lattice and figure (b) on the Husimi lattice. It is seen that the occupied-empty correlation is shorter-ranged than the directional correlations between occupied bonds.}
	\label{orientcorr}
\end{figure*}

Despite the differences between the atomistic simulations and our lattice models, we note the concentration of $12$, $21$, $23$ and $32$ water molecules in room temperature water is much smaller compared to the canonical 2-in-2-out molecules. Furthermore, as we will see shortly, the directional correlations within short loops that we discuss in the next section, are not so sensitive to the presence of defects. In addition, the variation of macroscopic properties across the phase diagram are mostly dominated by the correlations between the $22$ and other types of defects, which are well-captured by our model.\\

Besides the correlations between water molecules, it is also interesting to examine the behavior of the
empty sites with each other as well as with water molecules. In particular, we can construct the radial
distribution function between these empty sites which are essentially cavities/voids in a real network
with other sites in the lattice. Figure~\ref{voidgr} shows the pair correlation function associated
with these cavities. Perhaps not suprisingly, we see some manifestation of the hydrophobic effect in the
simple lattice model since the empty sites have a tendency to cluster with each other to form larger
cavities. We also observe that there is a preference for undercoordinated 12 and 11 water molecules to
cluster closer to the cavities.

\subsection{Orientational Correlations on the Lattice}

Besides the structural correlations, it is well known that orientational correlations in water can be quite long range on the order of 1-2nm. In fact, recent atomistic simulations by Galli and co-workers examined the orientational correlations in water (such as dipolar correlations) and showed that these are longer-ranged than the density-density correlations\cite{zhang2014dipolar}. We can examine the directional correlations of the H-bonds on the Bethe and Husimi lattices using a transfer matrix approach\cite{izmailian1998exact}. A transfer matrix $M$ is defined through the formula
\beq
M(x_n,x_{n+1}) = P(x_{n+1}|x_n)
\eeq
That is, the matrix element $M(x,y)$ gives the probability that a hanging bond at level $n+1$ is in state $y$ given that the bond at level $n$ is constrained to be in state $x$. $P(x|y)$ is calculated by constraining the two bonds to be in the required states and averaging over the other hanging bonds which can be done through the knowledge of the restricted partition functions.\\

Higher powers of $M$ then give
\beq
M^k(a,b) = P(x_{n+k}=b|x_n=a)
\eeq
That is, they give the probability of finding a bond in state $b$ at a distance $k$ away from a bond in state $a$. For large $k$, the probability $P(x_{n+k}=b|x_n=a)$ approaches the value $P(b)$, since the dependence on the state at level $n$ dies off.\\

We study the correlations between bonds separated by straight lines on the Bethe and Husimi lattice, since these show the strongest correlations. The bond-bond correlations are of two kinds: orientational and non-orientational. Orientational correlations manifest as correlations between the directionalities of two H-bonds separated by a distance $k$, that is, their propensity to be parallel or anti-parallel. Non-orientational correlations (related to density-density correlations) neglect the directionality of the hydrogen bonds and show up as the correlations between the occupation states of two bonds in the network that could either be empty or both be occupied.\\

\begin{figure*}[t]
	\centering
	\subfigure[]{
		\includegraphics[width=0.9\columnwidth]{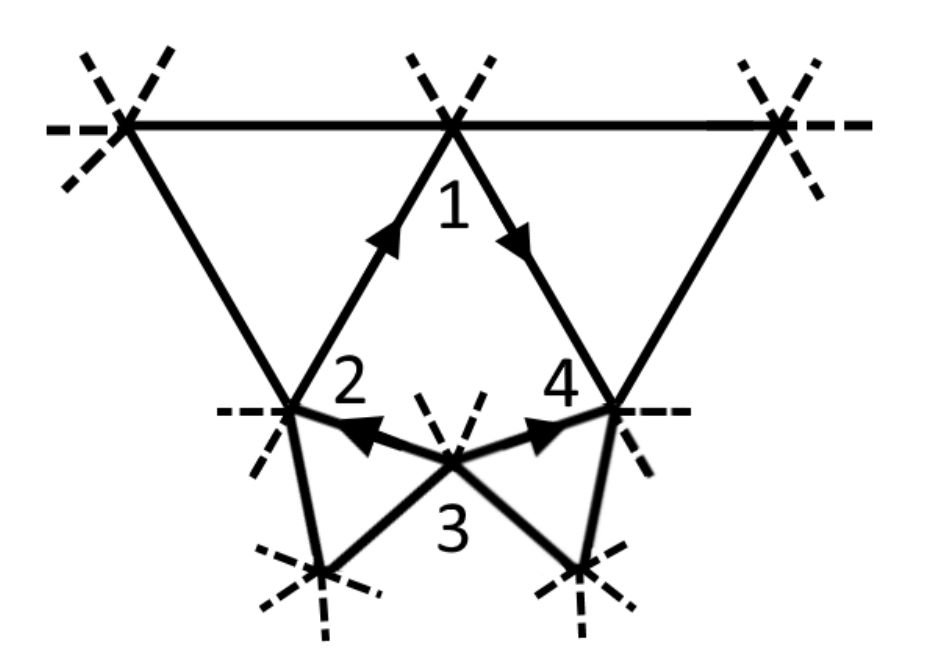}
	}
	\subfigure[]{
		\includegraphics[width=0.8\columnwidth]{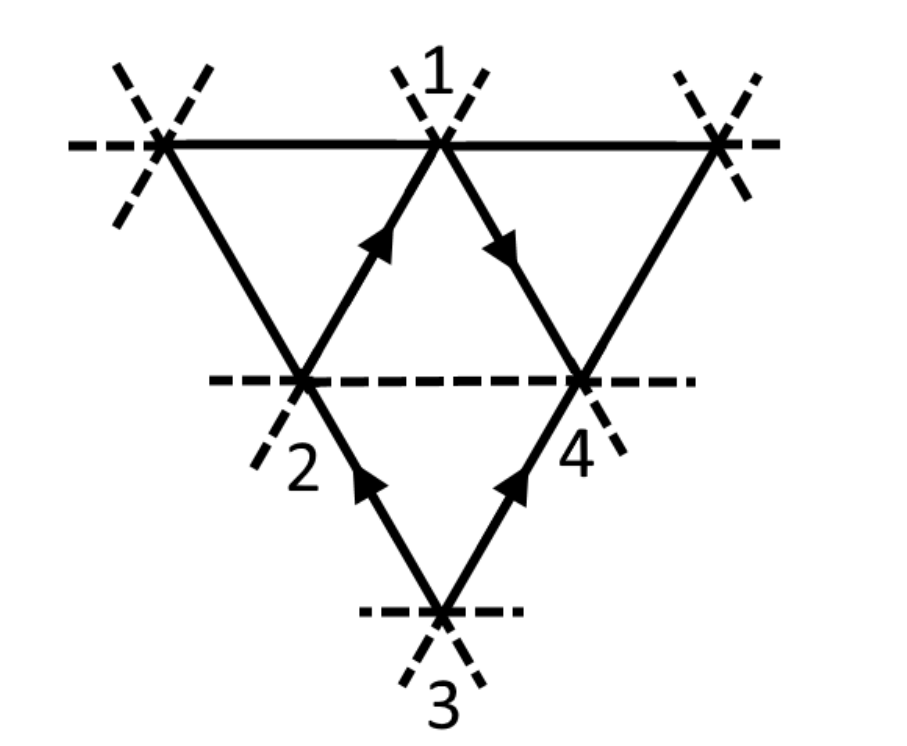}
	}
	\caption{Two kinds of 4-loops (of type 421) constructed from elementary triangles on the Husimi network: (a) an open loop (b) a loop with one interstitial }
	\label{4loops}
\end{figure*}

Fig~\ref{orientcorr} shows $P(x_{k}|x_{0})/P(x_k)$ on both the Bethe and Husimi lattices where the bond $x_o$ is constrained to be occupied. This quantity goes to $1$ as $k$ becomes large, as explained above. It can be seen from both the figures that the orientational correlations are longer-ranged than the non-orientational correlations. This is especially so on the Husimi lattice, which is to be expected as a larger number of loops on the lattice allows orientational correlations to propagate. The properties of the loops will be discussed next.

\subsection{Properties of Loops/Rings}

An important feature of the topology of the H-bond network in water is the distribution of the various types of closed rings or loops\cite{rahmanstillinger1973,bergman2000topological,hassanali2013proton} . The closed rings are also characterized by specific directional correlations which has important implications on the underlying topology of the network. There have been several studies that have examined the properties of closed rings and wires in the context of studying the properties of the bulk water\cite{rahmanstillinger1973}, proton transfer in water\cite{hassanali2013proton} and also hydrogen bond networks around biological systems\cite{rahaman2017,jong2017hydrogen}. We now examine some network properties of loops on the Bethe and Husimi lattices. \\

One can calculate various properties of loops on the Bethe and Husimi lattices despite the pure Bethe lattice not having loops of any kind, and the Husimi lattice not having loops of size $>3$. One does this, as described below, by deforming the lattice by hand. By doing this one can study properties of loops of a given size, but one cannot calculate, say, the distribution of sizes of loops, as the loops are constructed by hand.\\

The calculation of the properties of a loop with a particular size in the lattice models is done as follows. One deforms the lattice locally to have the structure of the desired loop, as shown in Fig. 7 a). In this case, the triangle on the original Husimi lattice shown in Fig 4 b) is distorted to create a loop of length 4. There are various other ways a loop can be constructed on the lattice. Fig~\ref{4loops} (b) shows another 4-loop, where it is required that vertices 2 and 4 should be within an interstitial distance of each other but not bonded. This loop is thus more compressed than the one in Fig~\ref{4loops} (a). Hence, adding more interstitials is equivalent to considering a more compressed geometry of the loop. On our lattice, open loops are those that do not contain any interstitial interactions. Under the assumptions that these loops are rare in the thermodynamic limit, one can calculate the weights of various loop configurations by summing over the configurations of the rest of the lattice consistent with the loop, using fixed points of the recursion relations on the appropriate mean-field network. \\

Bergman conducted molecular dynamics simulations of liquid water and used them to examine some interesting topological properties such as rings and specific hydrogen bond patterns within them\cite{bergman2000topological}. Following this work, we use the notation $ldm$ to classify the various types of loops. $l$ denotes the length of the loop, $d$ denotes the absolute difference between the number of anticlockwise and clockwise H-bonds along the loop, and $m$ denotes the number of vertices on the loop where anticlockwise and clockwise bonds meet. The loops shown in Figure \ref{4loops} (a) and (b) are thus of type $421$. Bergman calculated the frequencies of loops for all values of $d$ and $m$ for $l=4$ to $9$. Here we compare our results for $l=4$ and $l=6$ with Bergman's and show the analysis of some other loops in the SI \cite{supplinfo}.\\

Table \ref{4looptable} compares the proportions of the various types of $l=4$ loops calculated for open loops on the Bethe and Husimi lattices and 1-interstitial loops on the Husimi with the atomistic simulations of Bergman. It can be seen that the Bethe lattice, which is the simplest entropic model, does not capture the frequencies of the 4-loops very well. The best match is with open loops on the Husimi lattice, which signals that the triangular Husimi lattice captures loop structures better than a simple entropic model. Also, the fact that open loops, for types 421 and 440, perform better than 1-interstitial loops suggests that 4-loops of these types in room temperature water have an open structure.\\

\begin{table*}[]
	\centering
	\caption{Loop frequencies for 4-loops}
	\label{4looptable}
	\begin{tabular}{|l|l|l|l|l|l|}
		\hline
		Type of loop & Bergman &Bethe open & Husimi open & Husimi 1-i & Husimi open (only 22s) \\
		\hline
		401 & 0.17 & 0.20 & 0.081 & 0.10 & 0.08\\ 
		402 & 0.12 & 0.37 & 0.25 & 0.43 & 0.25 \\ 
		421 & 0.50 & 0.40 & 0.5 & 0.37 & 0.5 \\ 
		440 & 0.21 & 0.028 & 0.17 & 0.094 & 0.17\\ 
		\hline
	\end{tabular}
\end{table*}

\begin{table*}[]
	\centering
	\caption{Loop frequencies for 6-loops}
	\label{6looptable}
	\begin{tabular}{|l|l|l|l|l|l|l|}
		\hline
		Type of loop & Bergman & Bethe open & Bethe 1-i & Husimi open & Husimi 2-i & Husimi 1-i (only 22s)\\
		\hline
		601 & 0.12 & 0.217 & 0.219 & 0.102 & 0.130 & 0.129\\ 
		602 & 0.089 & 0.122 & 0.114 & 0.204 & 0.073 & 0.068\\ 
		603 & 0.005 & 0.005 & 0.005 & 0.023 & 0.003 & 0.002\\ 
		621 & 0.25 & 0.219 & 0.219 & 0.204 & 0.263 & 0.266\\ 
		622 & 0.13 & 0.0855 & 0.086 & 0.273 & 0.110 & 0.098\\ 
		641 & 0.26 & 0.219 & 0.219 & 0.171 & 0.263 & 0.262\\ 
		660 & 0.14 & 0.14 & 0.14 & 0.023 & 0.157 & 0.173\\
		\hline
	\end{tabular}
\end{table*}

\begin{table*}[]
	\centering
	\caption{Loop frequencies for 6-loops at different Temperatures and Pressures}
	\label{6looptable2}
	\begin{tabular}{|l|l|l|l|}
		\hline
		Type of loop & $T=1$, $P'=0$ & $T=0.6$, $P'=0$ & $T=1$, $P'=0.5$ \\
		\hline
		601 & 0.130 & 0.130 & 0.130 \\ 
		602 & 0.073 & 0.074 & 0.073 \\ 
		603 & 0.003 & 0.003 & 0.003 \\ 
		621 & 0.263 & 0.264 & 0.262 \\
		622 & 0.110 & 0.109 & 0.110\\ 
		641 & 0.263 & 0.262 & 0.264 \\ 
		660 & 0.157 & 0.156 & 0.157 \\
		\hline
	\end{tabular}
\end{table*}

Atomistic based simulations have shown that both ice and liquid water are dominated by a large number of six-membered rings\cite{rahmanstillinger1973,hassanali2013proton}. We thus also examined the properties of the 6 loops on both the Bethe and Husimi lattice. Table \ref{6looptable} compares the loop statistics for $l=6$ loops. Again, it is seen that the best prediction comes from the Husimi lattice whereas the Bethe lattice does not capture the qualitative trends in loop populations. Comparing the open loops to those with 2 interstitials, we find that the latter reproduces the properties of the loops more consistently compared to the atomistic simulations. Thus unlike the 4-loops, our analysis suggests that the 6-loops with more interstitials are more likely to be in a compressed geometry.\\

It is also interesting to examine whether the network properties of the loop are sensitive to the presence of defects. The last column of Tables~\ref{4looptable} and ~\ref{6looptable} shows the open Husimi lattice computed for type $22$ molecules only. It is seen that this does not affect the frequencies by much compared to the Husimi open lattice with defects. This thus suggests that the $22$ molecules are sufficient to obtain the loop frequencies and that the other types of molecules play a smaller part in reproducing 
the directional correlations within the loops.\\

The preceding analysis focuses on the properties of loops at ambient temperature and pressure. Varying the weights with the T and P as described earlier, we also examined the variations in properties of the loops as a function of temperature and pressure. Table~\ref{6looptable2} shows the fractions of various types of 6-loops with two interstitials on the Husimi lattice at three different values of $(P,T)$. Within the liquid phase, the proportions of the loops do not seem to change significantly with pressure and temperature. One explanation of this is that, as described in the previous paragraph, that the proportions of 6-loops are mainly governed by 22 molecules which does not change significantly with temperature and pressure.

Besides the simulations of Bergman, a recent molecular dynamics study by Hassanali and co-workers workers was used to examine the mechanisms of proton transfer in liquid water\cite{hassanali2013proton}. They found that water molecules are threaded by closed rings with specific directional correlations between them. They classified the water molecules into three types: DD which donate two hydrogen bonds, AA which accept two hydrogen bonds and finally DA which accept and donate one hydrogen bond, always within the ring. In these studies it was found that most closed loops or rings in the simulations were dominated by a single DD-AA pair. Within the context of our analysis here and comparing with that of Bergmans, these loops correspond to those with the index $m$=1. Since our lattice models are also dominated by loops with a single DD-AA pair, this is consistent with the simulation data. This analysis shows that the existence of water wires in the hydrogen bond network can easily be rationalized using simple entropic models and do not require a sophisticated intermolecular potential.

\section{Anomalies of Water in Lattice Model} \label{sec:vari}

Up to this point in the manuscript we have shown that at normal temperatures there is a rich diversity of interesting network correlations between sites on the lattice and within longer-range structures like loops. These features are also present in atomistic simulations. We now move on to exploring the water anomalies that occur by varying temperature and pressure and also to understand the role of water defects in this regard.\\

\begin{figure*}[]
	\centering
	\subfigure[]{
		\includegraphics[width=0.8\columnwidth]{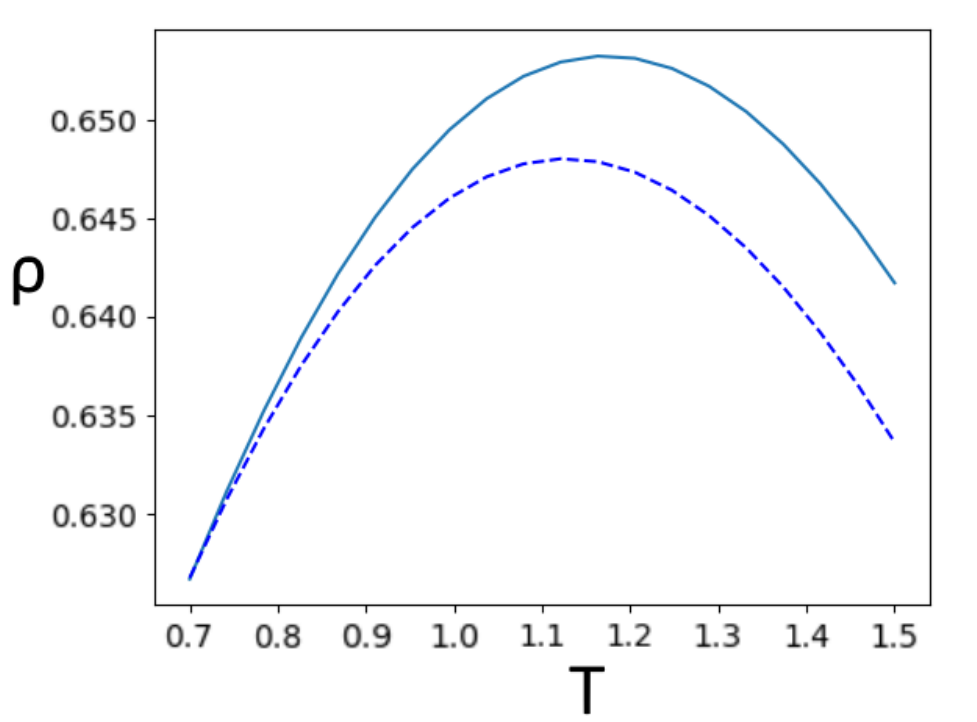}
	}
	\subfigure[]{
		\includegraphics[width=0.8\columnwidth]{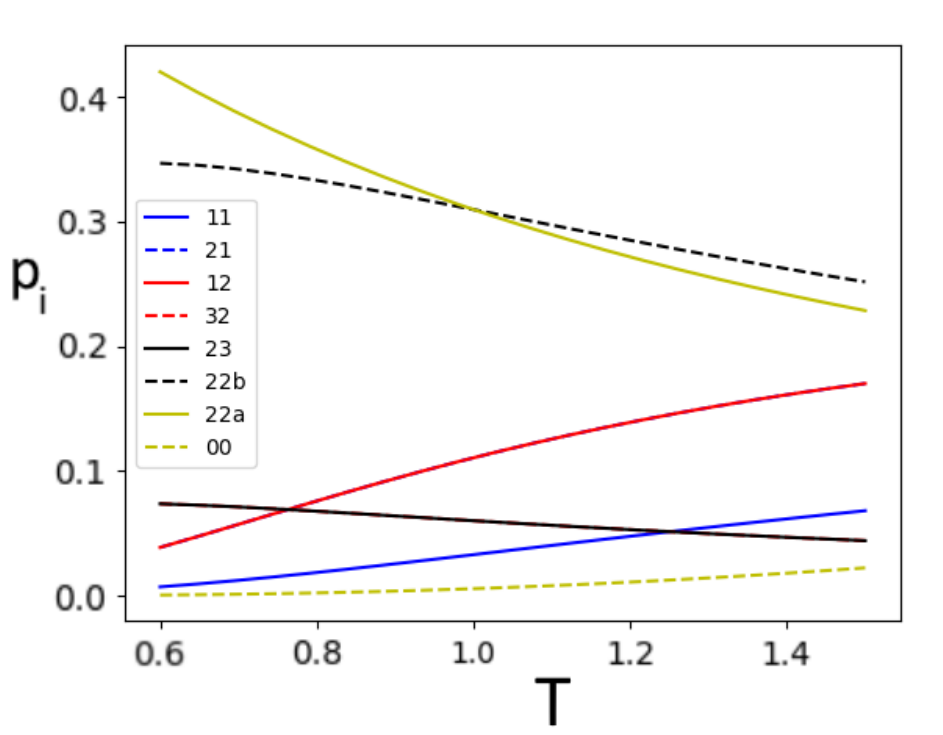}
	}
	\caption{(a) Density vs Temperature (solid line) and Density vs Temperature when the defects are assigned the same value as the non-tetrahedral 22s (dashed line) and (b) the variation of the concentrations of various molecule types with temperature}
	\label{den}
\end{figure*}

We begin by first examining the variation of density as a function of T predicted by our lattice model. In order to compute the density, we assume that the total volume of the liquid is given by a sum of local volumes. If the fraction of voids is denoted by $f_{void}$, the density is given by
\beq
\rho(T,P) = \frac{N(T,P)}{V(T,P)} = \frac{1-f_{void}}{\sum_i v_i}
\eeq

\noindent where the numerator results from the fact that the voids don't count towards the total number of molecules, but they do count towards the total volume, ie, in the denominator. Recall that $v_i$ are the local volumes which are given by Equation~\ref{eq:vols}.\\

Figure~\ref{den} a) shows the variation of density as a function of temperature where we observe the density maximum at around $T\sim1.1$. The right panel of Figure~\ref{den} b) shows the variation in the concentration of different types of water molecules in the hydrogen bond network of our lattice model. The existence of a density maximum can be attributed to the fact at lower temperatures the number of tetrahedral 22 molecules which have a higher local volume, increases, while the increase in volume at higher temperature is attributed to the increased number of voids and defects \cite{sastry1996singularity}. \\

Our lattice model with defects allows us to examine the origin of the density maximum in terms of the local coordination of different sites. The non-22 molecules included in our model have a different volume from the non-tetrahedral 22s, and a different variation with temperature. Thus the total volume of the non-22 molecules has a different temperature dependence from the 22s. In order to assess the importance of the inclusion of the non-22 molecules, we also show in Fig~\ref{den} a), in dashed blue, the variation of density as a function of temperature where all the non-22s were assigned the same volume as that of non-tetrahedral 22 molecules (the local volumes of tetrahedral 22s and the voids are still different from these, and hence we still see a density maximum). We see here that there while there is still a density maximum, the curve has shifted quite significantly. Thus the location of the density maximum is sensitive to the specific assignment of the volume of different defects.\\

One of the other interesting anomalies of water is the existence of a compressibility minimum as a function of temperature \cite{stanley1998puzzling}. The compressibility for water shows an anomalous behaviour as compared to a regular liquid since it rises at low temperatures. The compressibility is the susceptibility with respect to pressure and is defined as the following derivative
\beq
\kappa = \frac{1}{V} \left( \frac{dP}{dV} \right)_{T,N}
\eeq

\begin{figure}[]
	\centering
	%\subfigure[]{
		\includegraphics[width=0.8\columnwidth]{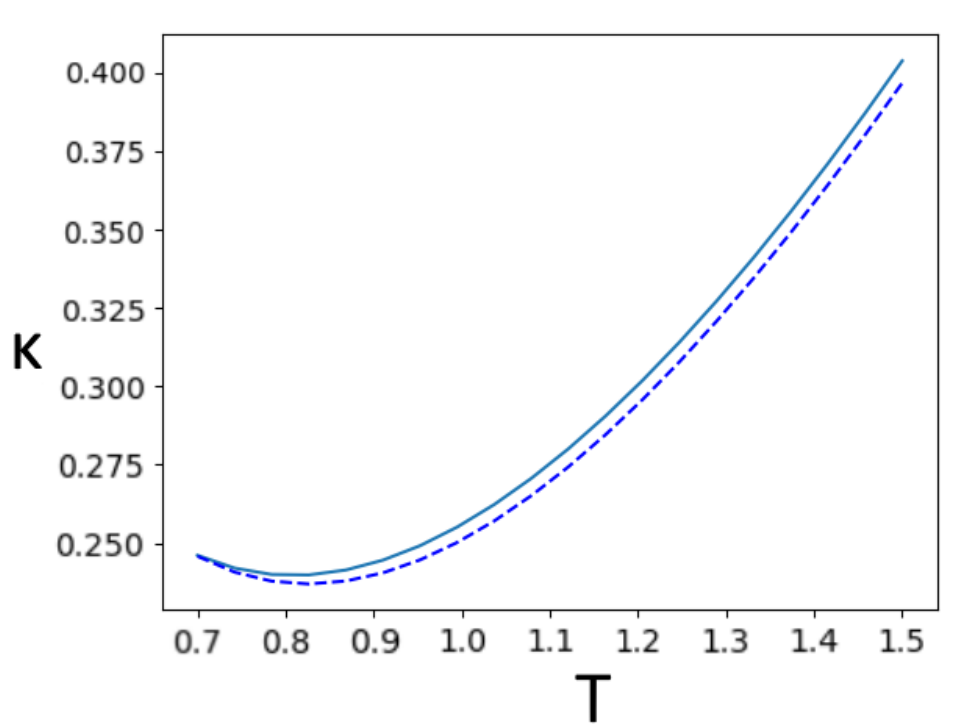}
	%}
	%\subfigure[]{
	%	\includegraphics[width=0.8\columnwidth]{corrvsT.pdf}
	%}
	\caption{Compressibility vs Temperature (solid line) and Compressibility vs Temperature when the defects are assigned the same value as the non-tetrahedral 22s (dashed line)}
	\label{compr}
\end{figure}

Figure~\ref{compr} shows the compressibility against temperature. The compressibility in our lattice model shows anomalous behaviour, rising at low temperatures. Unlike the density maximum, the position of the compressibility minimum is less sensitive to the volume assigned to the defects - see dashed blue line in Figure~\ref{compr}. The physical origin of the compressibility minimum of water has been discussed in the literature with several interpretations. In the two-critical-point scenario, there is a second-order phase transition at some temperature $T_c<1$ and pressure $P_c$, between more tetrahedral and less tetrahedral phases. Thus, the correlations between tetrahedral regions diverge at the critical point, and the compressibility rise at low temperatures is a result of the such increasing correlations as the temperature is decreased. However, in our model, we find that the correlation length between tetrahedral 22s shows only a very weak increase as the temperature is lowered. This is inconsistent with the prediction of formation of tetrahedral patches as the temperature is lowered, but consistent with the phase diagram obtained in the next section, which does not show a second critical point in the supercooled region. \\

\begin{figure}[]
	\centering
	\includegraphics[width=0.8\columnwidth]{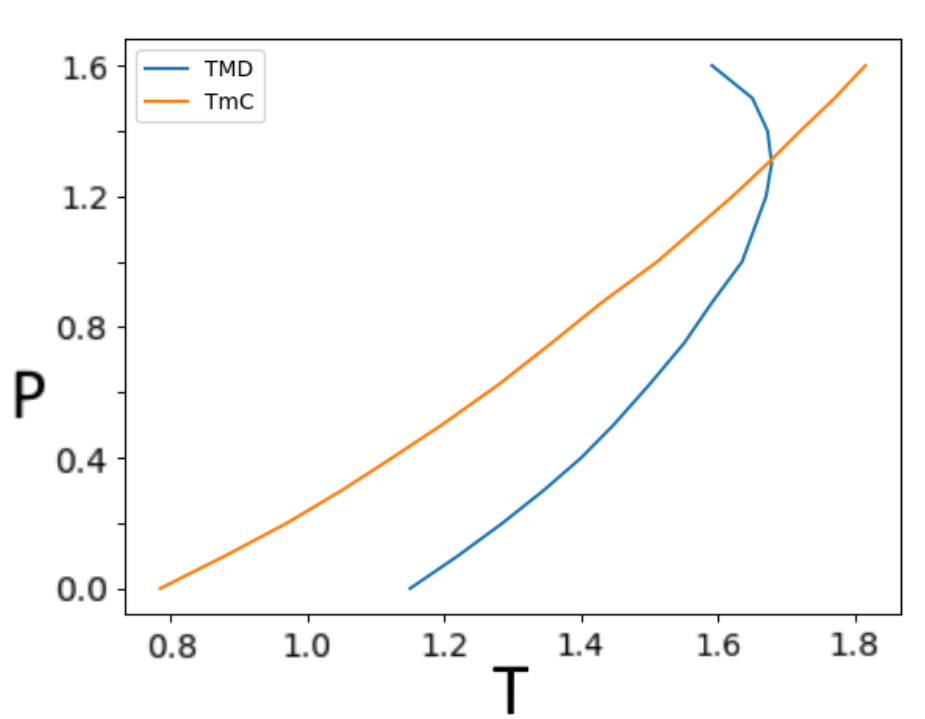}
	\caption{The lines of the Temperatures of Maximum density (TMD, blue or dark gray line) and the Temperatures of minimum compressibility (TmC, red or light gray line) in the P-T plane.} %The dashed lines denote the same curves when the defects are assigned the same value as the non-tetrahedral 22s.}
	\label{tmd}
\end{figure}

Another explanation for the compressibility anomaly was put forward by Sastry et al. \cite{sastry1996singularity} who argued that a negatively sloped Temperature of Maximum Density (TMD) line can cause a negatively sloped compressibility curve at low temperatures, even in the absence of a second critical point. To test whether our model exhibits behaviour consistent with this scenario, we show in Figure ~\ref{tmd} the TMD and Temperature of minimum Compressibility (TmC) lines in the P-T plane. At atmospheric pressures, the TMD is positively sloped. At high pressures above $P \approx 1.4$, the two lines do indeed cross and the TMD becomes negatively sloped, as required by thermodynamic consistency. As we report in the next section, we find that our models exhibit a phase diagram without a liquid-liquid critical point. The behaviour of the TMD and TmC lines reinforces the point that the TmC can arise even in the absence of a liquid-liquid critical point. It should be noted here that recent work has indicated that the increase in compressibility seen in singularity-free models on lowering temperature does not seem to be sharp enough to explain the experimental data on water \cite{pathak2016structural}.\\

\section{Phase Diagram of the Network} \label{sec:phase}

\begin{figure}[b]
	\centering
	\subfigure[]{
		\includegraphics[width=0.8\columnwidth]{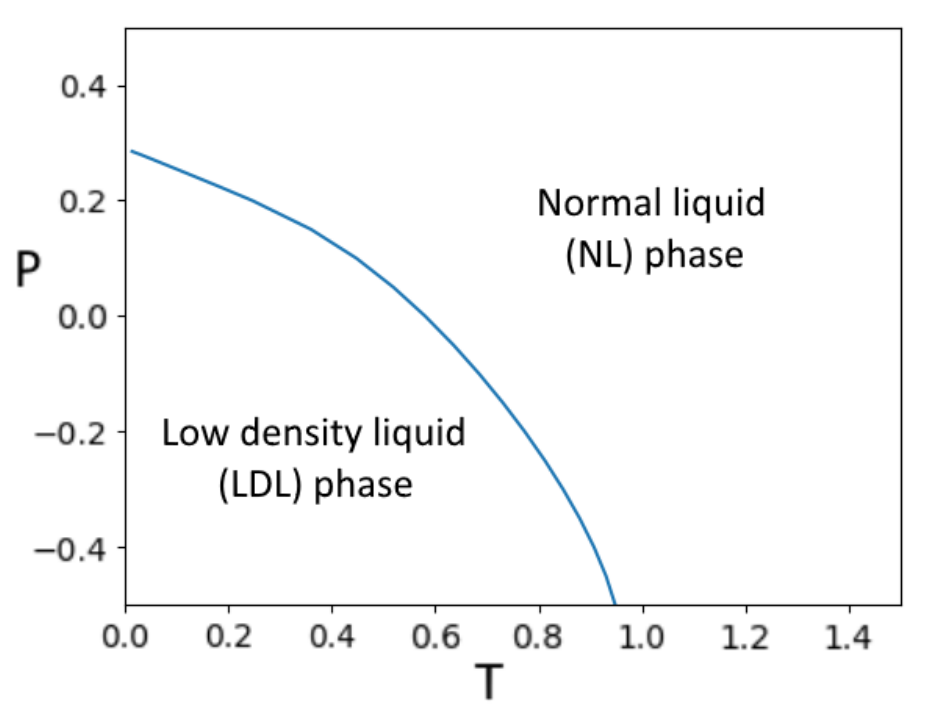}
		
	}
	\hspace{0.5cm}
	\subfigure[]{
		\includegraphics[width=0.8\columnwidth]{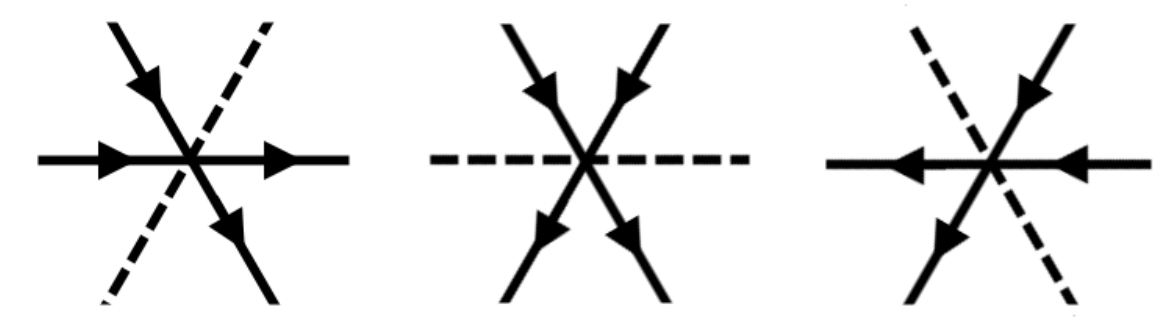}
	}
	\caption{(a) The phase diagram calculated on the Bethe Lattice network. The blue line denotes the line of first order phase transitions, (b) The three different orientations of tetrahedral 22s, which correspond to the three symmetry-broken LDL phases}
	\label{pdbethe}
\end{figure}

We now discuss the phase diagram that we obtain in our model. The phase transition lines are calculated by determining the fixed points of the recursions discussed in Section 2, and finding the crossing points of the free energies associated with the two phases. The phase diagram for our model on the Bethe Network is shown in Figure~\ref{pdbethe}. It shows two phases with a symmetry breaking transition between them. In keeping with terminology in the literature and the fact that tetrahedral molecules have higher local volumes, we call the tetrahedral phase a `low density liquid' (LDL) and the liquid phase the `normal liquid' which has a higher density (NL).\\

The liquid-gas transition is not present in our Bethe lattice model because we only take into account a few kinds of defects, which are relevant at room temperatures and below. As the temperature is increased, defects with lower bonding, become relevant to the physics. On the Bethe lattice, molecules with zero bonds or one bond would be important for a liquid-gas transition, as these allow the formation of small isolated clusters of water on this lattice. However, this is beyond the range of our present study.\\ %Perhaps on the Husimi lattice, formation of triangular clusters of three mutually bonded 11s might give one a liquid-gas transition even without including extra types of molecules, but if this happens it is at temperatures beyond the range that we consider in this study.\\

The free energy change associated with this crossing has a first-order discontinuity and hence the LDL and NL phases are separated by a first order phase transition. Looking at the densities of various molecule types across the phase transition, we see that the density of all molecule types changes discontinuously across the transition, which confirms the first-order nature of the transition. We find that the phase transition is everywhere first-order, and there is no critical point. \\

The high temperature phase breaks no symmetries, and is characterized in detail in the next section. At low temperatures and low or moderate pressures there is a first-order transition to a tetrahedral liquid phase in which almost all the molecules are of the tetrahedral 22 type. There are three possible tetrahedral orientations of the 22s which are shown
in the bottom panel of Figure~\ref{pdbethe}. Subsequently, there are three possible orientations of the symmetry-broken phase. The model on the Husimi network also shows only a first order phase transition.\\

Similar models have been studied in the past, with some reaching different conclusions about the structure of the phase diagram. The model studied by Sastry and co-workers \cite{sastry1996singularity}, mentioned in the Introduction, has directional hydrogen bonds with configurations weighted by the number of hydrogen bonds but no preference for tetrahedrality in the local geometry. They do not observe a phase transition and furthermore do not find a critical point. Sastry et al's model is similar to be our model with $e=1$, where we have distortion but no preference for tetrahedrality. Setting $e=1$ in our model, we also find a phase diagram without an LDL phase. Thus, with distortion but without a preference for tetrahedrality, there is no phase transition. We can also study our model without distortion, setting $b=c=d=0$, and study the effect of tetrahedrality alone. In this case we find that there is a first-order phase transition between an NL phase for low $e$ and an LDL phase for high $e$.\\

It is thus evident that the phase transition is caused as an effect of co-operativity between H-bonds on the same molecule (our interaction $e$). This interaction has been included in two different ways in the literature on mean-field lattice models of water. Franzese and Stanley\cite{franzese2002liquid} modified the model of Sastry discussed above, to add a external field that is proportional to the density of particles in a broken-symmetry state, and whose strength is interpreted as the co-operativity interaction. They found a phase diagram with a liquid-liquid critical point, in the case when the cooperative interaction is much smaller than all the other interactions in the model, in agreement with Monte-Carlo simulations of a related model\cite{stokely2010effect}.\\
	
Heckmann and Drossel \cite{heckmann2014}, by contrast, add the co-operative term as an internal interaction in the Hamiltonian, which is solved by the Curie-Weiss mean-field method. This results in a first-order liquid-liquid transition without a critical end-point. The same result was obtained by Franzese et al. under a similar approximation \cite{franzese2000hydrogen,franzese2002theory}. As our  Bethe lattice treatment includes the co-operativity effect as an internal coupling, the phase diagram we obtain, shown in Fig. \ref{pdbethe}, is consistent with the results of Heckmann and Drossel.

\section{Conclusions}

In this work, we have made first steps in the development of a lattice
model of the hydrogen bond network in water, aimed at making a stronger connection with
information obtained from atomistic simulations. We present two
lattice models that are analogous to graph-theoretical models used in
the literature for applications in the understanding of
networks. Specifically, our water networks are built as vertex models, and studied on the
6-coordinated Bethe and Husimi lattices which have been extensively
used in the statistical physics community. \\

The input for our model is inspired by recent atomistic simulations of
water showing that two-thirds of water are have 2 incoming and 2 outgoing hydrogen bonds, while
the rest consists of local coordination
defects\cite{gasparotto2016probing}.  Tetrahedral waters and defects
have a tendency to cluster with each other in the network.  Our models
include a description of not only tetrahedral water
molecules, but also the important coordination defects. Comparing the
pair-correlation functions on the lattice to those obtained from the
atomistic simulations, we show that our simple models are able to
capture many of the important physics of the correlations  associated
with the tetrahedral waters. This demonstrates that the clustering
with respect to the most dominant waters in the network is driven
purely by entropic effects. \\

Besides examining local coordination defects and the correlations
between them and tetrahedral waters, we also studied longer-range
correlations associated with the hydrogen bond patterns in
closed loops. These types of directional correlations play an
important role in understanding the statistical properties
of water or proton wires in the hydrogen bond network. Interestingly,
we can show that the directional correlations within closed loops
predicted from some of our models are also consistent with results
from previous atomistic simulations. Furthermore, we also show that
for our lattice models, these types of network correlations
are mostly reproduced by the presence of tetrahedral water
molecules. \\

The phase diagram of our lattice model reinforces the singularity free scenario proposed by Sastry and co-workers - we do not find any existence of a critical point. The fact that this emerges
from a model where the microscopics has been adequately built in, gives more confidence to this observation. In addition, our model also captures some of the anomalies of water, although, as mentioned in the previous section, the increase in compressibility seen in singularity-free models might not be sharp enough to explain the observational data on water \cite{pathak2016structural}.\\

As alluded to earlier in the introduction, part of our motivation
of developing this model is to delve deeper into the dynamics
of the hydrogen bond network of water. In particular, our models
provide a good starting point for understanding the mechanisms
associated with network reorganization at equilibrium or approaching
equlibrium after a perturbation. There have also been various 
suggestions from atomistic simulations of liquid water of
long-range and timescales associated with network relaxation
involving structures such as rings or loops or wires
\cite{ohmine1992,ohmine1996,ohmine1999,
hassanali2013proton,jong2017hydrogen,
gibertihassanali2017}. Since dynamics is ultimately driven by the rare fluctuations,
the movement between different local coordination defects is
also likely to play an important role. Our lattice model
captures many of these microscopic details and work is currently
underway to describe dynamics on these network models.

%\section{Supplementary Material}

%In the supplementary information, we provide results for the loop frequencies for 5- and 7-loops, and for the $g(r)$ around all molecule types on the Bethe and Husimi lattices.\\

%We see that the results for the loops show that the Husimi lattice predicts the loop frequencies for 5- and 7-loops reasonably well. For the $g(r)$ around various types of molecules, one sees that the Husimi lattice does not improve the Bethe lattice predictions significantly, and shows the same shortcomings at predicting the $g(r)$ around molecules of types $21$, $12$, $32$ and $23$.

\section{Acknowledgements}

We wish to thank Deepak Dhar for useful comments on the manuscript.

\bibliographystyle{unsrt}
\bibliography{water}

\end{document}

\maketitle

In this supplementary information, we provide results for the loop frequencies for 5- and 7-loops, and for the $g(r)$ around all molecule types on the Bethe and Husimi lattices.\\

We see that the results for the loops show that the Husimi lattice predicts the loop frequencies for 5-loops reasonably well, even if one only includes loops with only molecules of type $22$ on the vertices. For the 7-loops, the Husimi loop with 2 interstitials is very accurate at predicting the results from simulations. Here too, restricting the loops to only have molecules of type $22$ does not significantly change the statistics.\\

For the $g(r)$ around various types of molecule, one sees that the Husimi lattice does not improve the Bethe lattice predictions significantly, and shows the same shortcomings at predicting the $g(r)$ around molecules of types $21$, $12$, $32$ and $23$.

\newpage

\section{Loop statistics for lengths 5 and 7}

\begin{table}[h!]
	\centering
	\caption{Loop frequencies for 5-loops (simulation data taken from Bergman \cite{bergman})}
	\label{5looptable}
	\begin{tabular}{|l|l|l|l|l|}
		\hline
		Type of loop & Bergman & Husimi 1-i & Husimi 2-i & Husimi 1-i (only 22s) \\
		\hline
		511 & 0.34 & 0.34 & 0.35 & 0.34\\ 
		512 & 0.11 & 0.16 & 0.17 & 0.17 \\ 
		531 & 0.35 & 0.35 & 0.35 & 0.35 \\ 
		550 & 0.20 & 0.15 & 0.14 & 0.15\\ 
		\hline
	\end{tabular}
\end{table}

\begin{table}[h!]
	\centering
	\caption{Loop frequencies for 7-loops (simulation data taken from Bergman \cite{bergman})}
	\label{7looptable}
	\begin{tabular}{|l|l|l|l|l|}
		\hline
		Type of loop & Bergman & Husimi 1-i & Husimi 2-i & Husimi 2-i (only 22s) \\
		\hline
		711 & 0.19 & 0.185 & 0.182 & 0.179\\ 
		712 & 0.18 & 0.19 & 0.182 & 0.183 \\ 
		713 & 0.017 & 0.021 & 0.018 & 0.018 \\ 
		731 & 0.19 & 0.186 & 0.193 & 0.193\\ 
		732 & 0.12 & 0.135 & 0.13 & 0.131\\ 
		751 & 0.20 & 0.20 & 0.21 & 0.21\\ 
		770 & 0.10 & 0.081 & 0.086 & 0.085\\ 
		\hline
	\end{tabular}
\end{table}

\newpage

\section{Bethe lattice radial distribution functions}

\begin{figure}[h!]
\centering
\subfigure[]{
	\includegraphics[width=0.46\textwidth]{bethe_22a_1.png}
}
\subfigure[]{
	\includegraphics[width=0.46\textwidth]{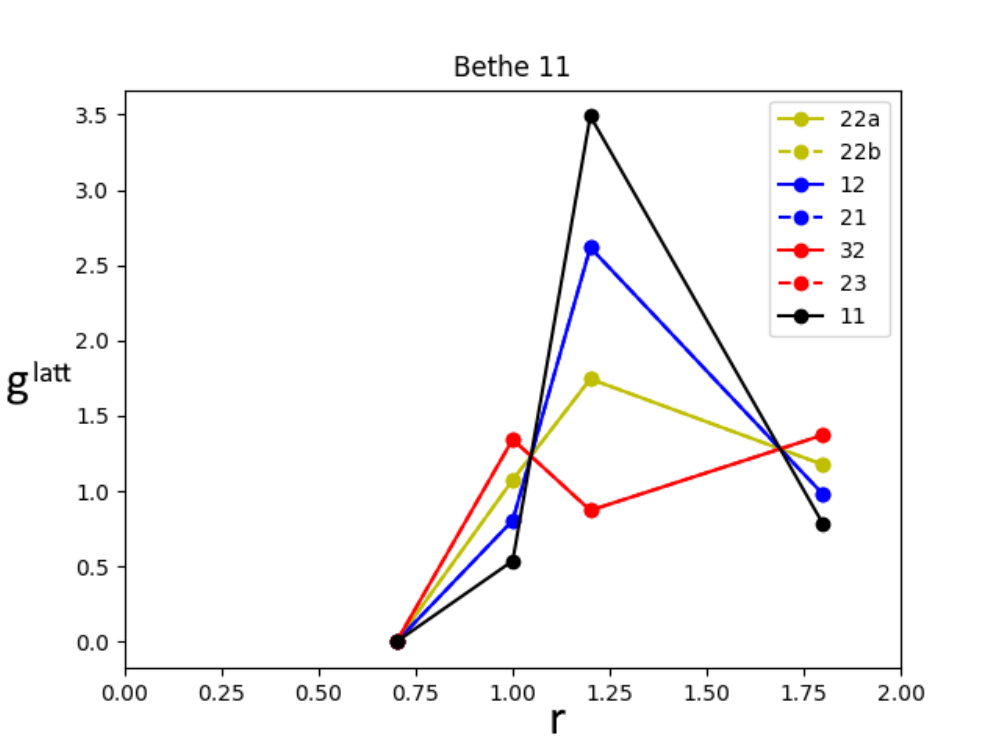}
}
\subfigure[]{
	\includegraphics[width=0.46\textwidth]{bethe_12_1.png}
}
\subfigure[]{
	\includegraphics[width=0.46\textwidth]{bethe_21_1.png}
}
\subfigure[]{
	\includegraphics[width=0.46\textwidth]{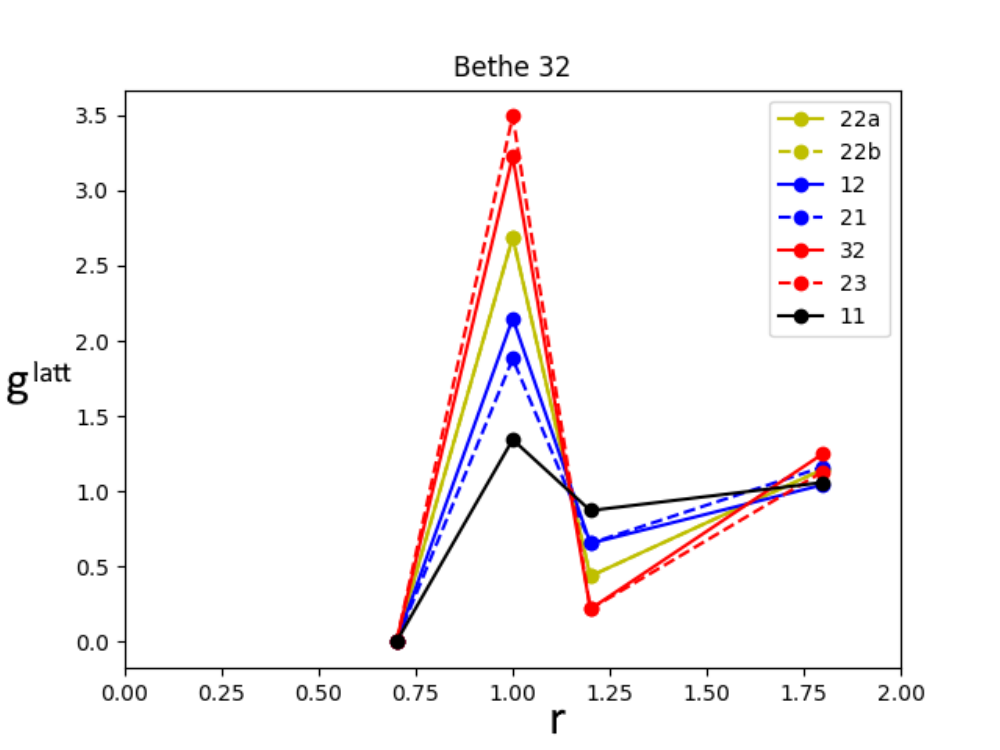}
}
\subfigure[]{
	\includegraphics[width=0.46\textwidth]{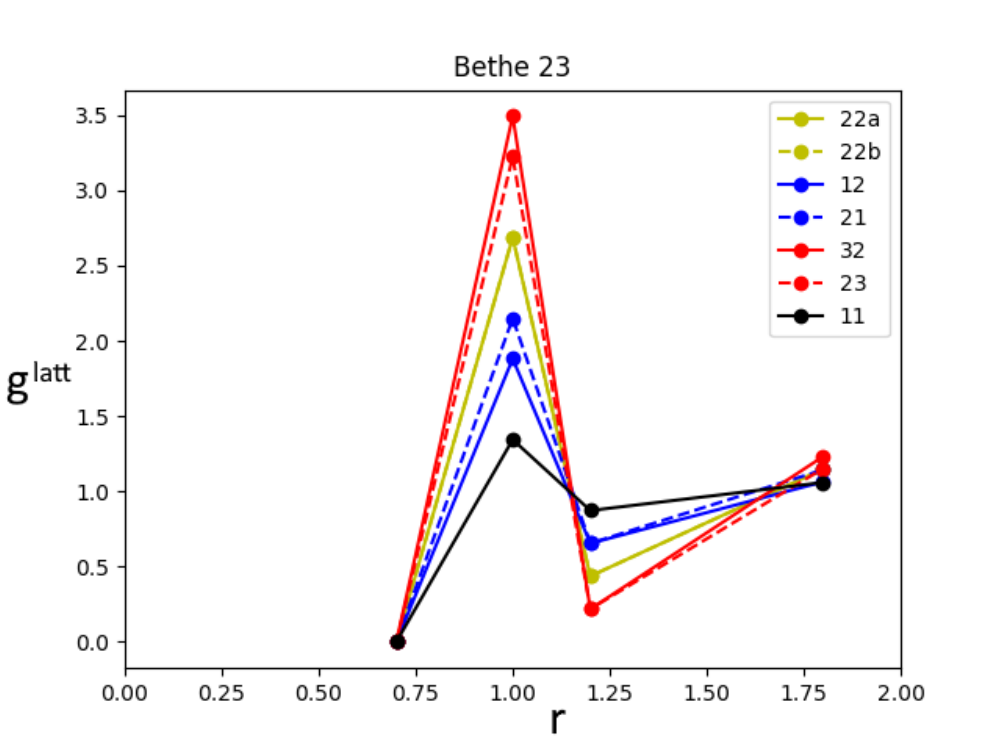}
}
\caption{The $g(r)$ around various types of molecules on the Bethe Lattice}
\label{gr_21}
\end{figure}

\newpage

\section{Husimi lattice radial distribution functions}

\begin{figure}[h!]
	\centering
	\subfigure[]{
		\includegraphics[width=0.46\textwidth]{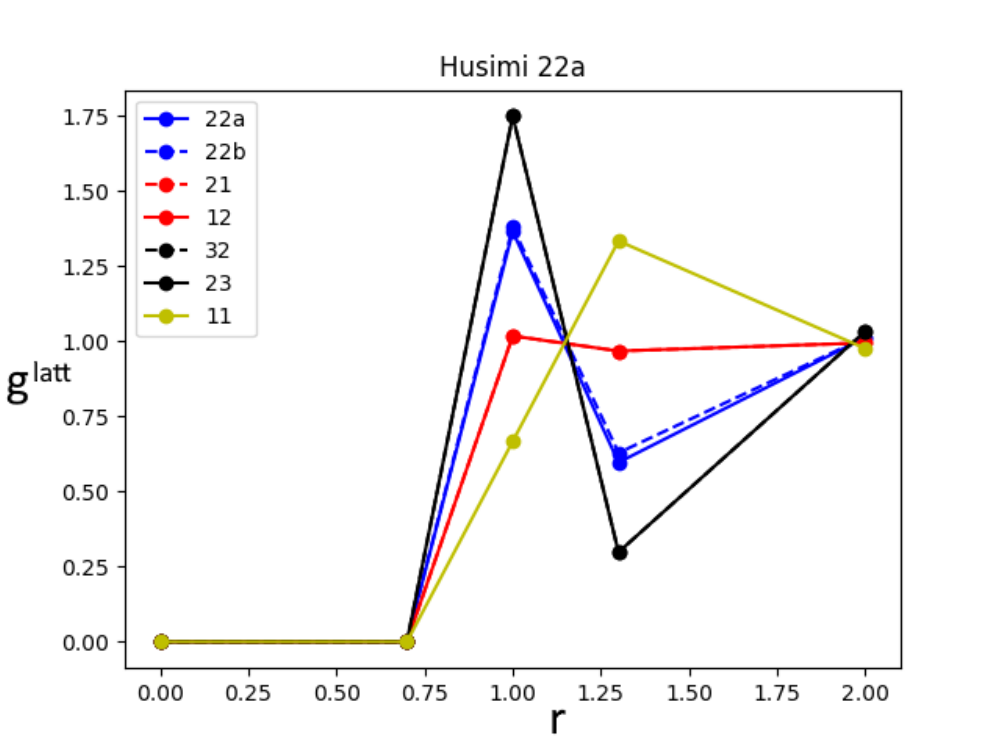}
	}
	\subfigure[]{
		\includegraphics[width=0.46\textwidth]{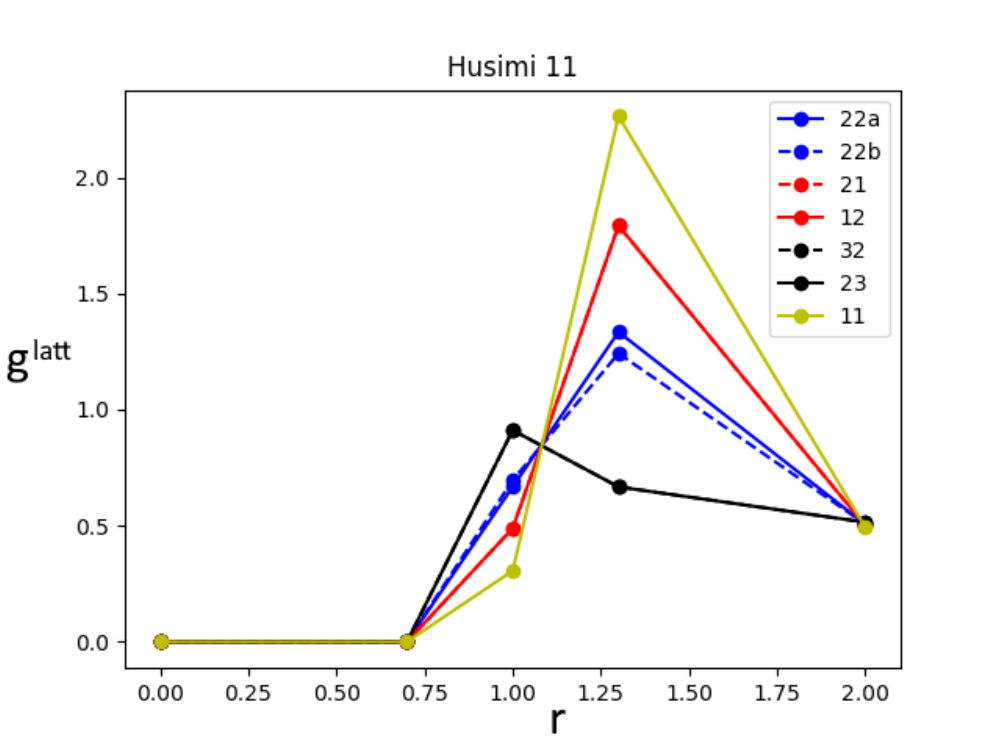}
	}
	\subfigure[]{
		\includegraphics[width=0.46\textwidth]{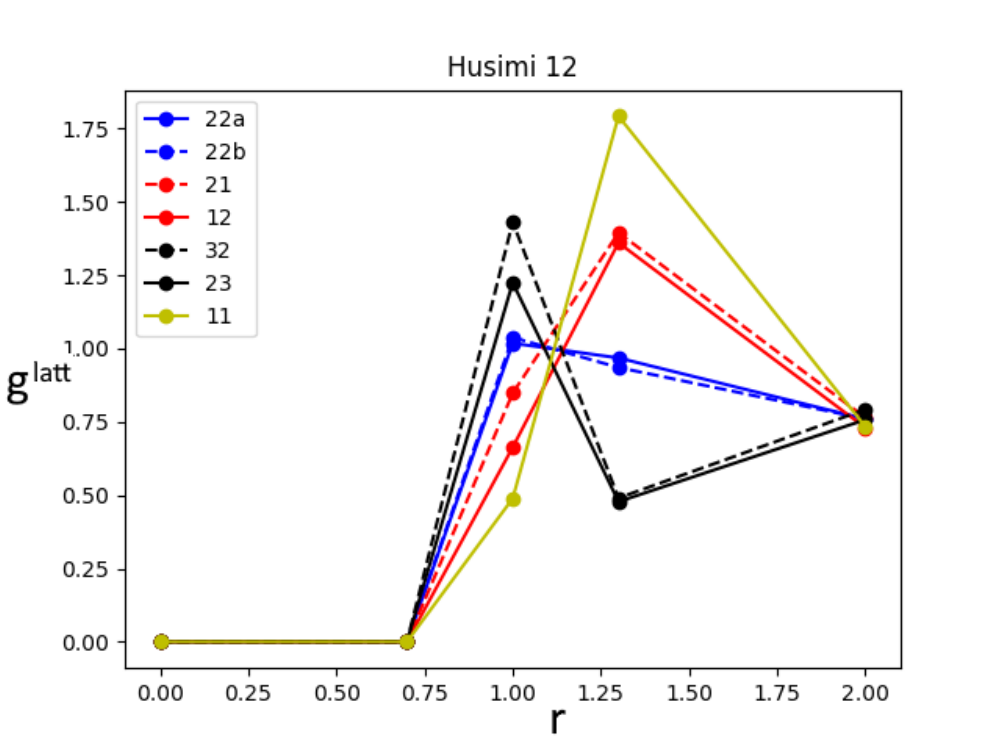}
	}
	\subfigure[]{
		\includegraphics[width=0.46\textwidth]{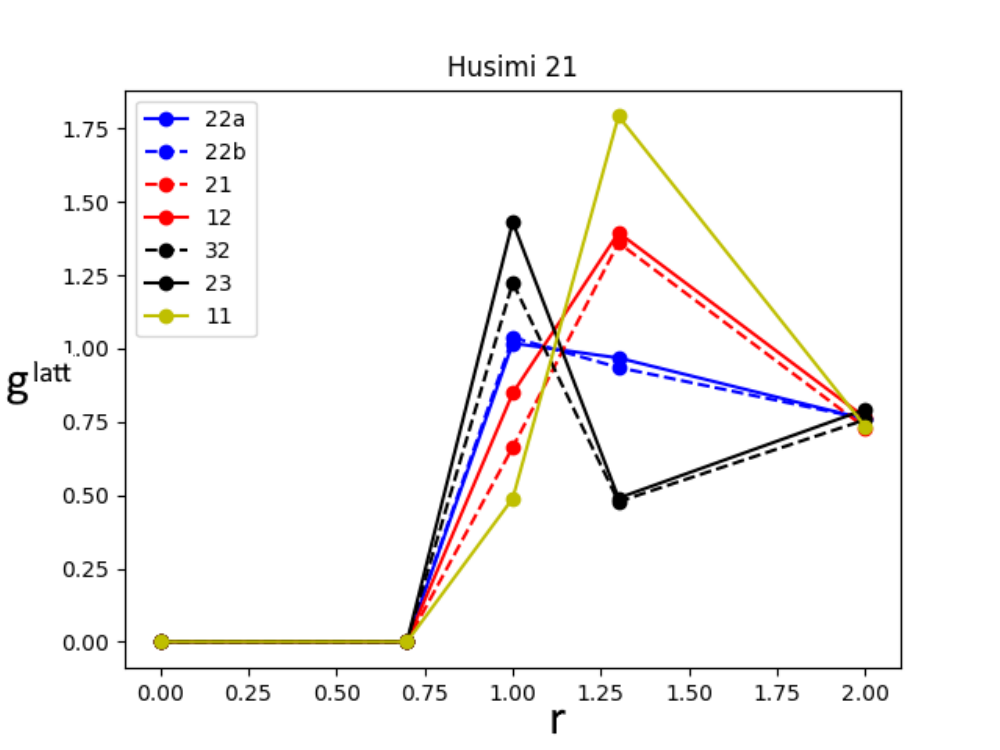}
	}
	\subfigure[]{
		\includegraphics[width=0.46\textwidth]{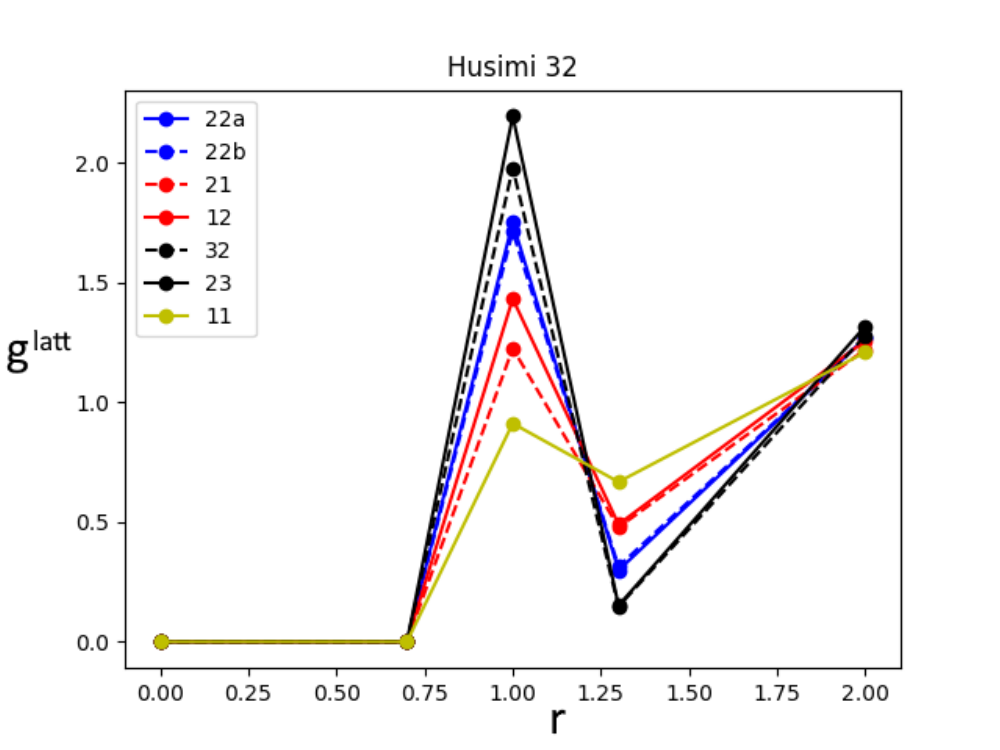}
	}
	\subfigure[]{
		\includegraphics[width=0.46\textwidth]{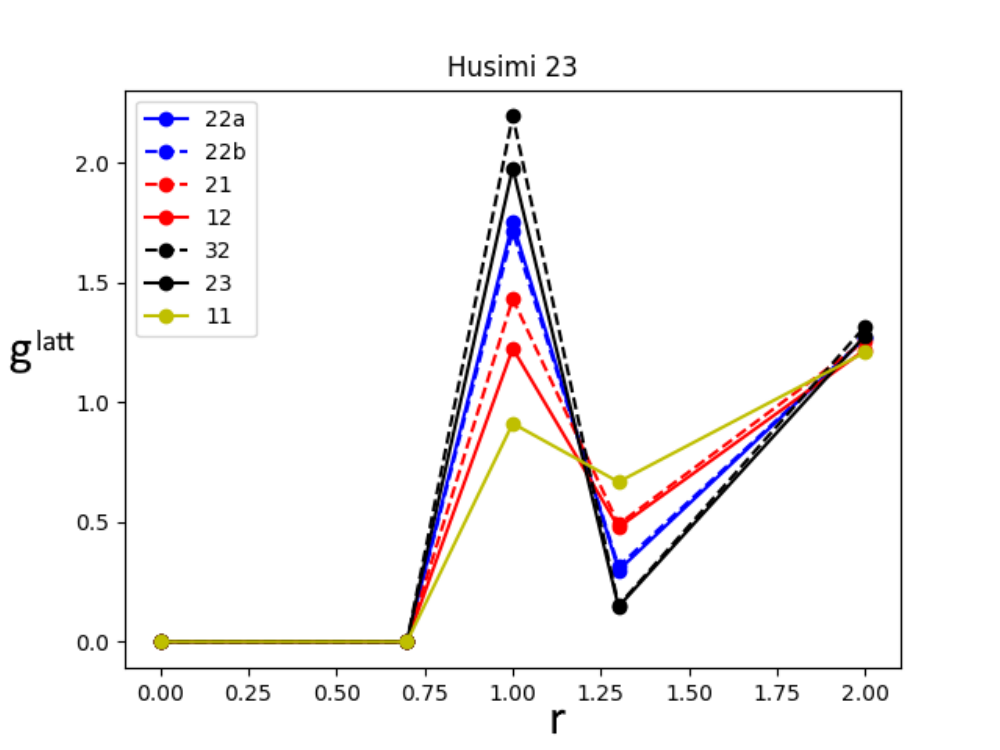}
	}
	\caption{The $g(r)$ around various types of molecules on the Husimi Lattice}
	\label{gr_21}
\end{figure}